\documentclass[pdflatex,sn-mathphys-num]{sn-jnl}

\usepackage{graphicx}%
\usepackage{multirow}%
\usepackage{amsmath,amssymb,amsfonts}%
\usepackage{amsthm}%
\usepackage{mathrsfs}%
\usepackage[title]{appendix}%
\usepackage{xcolor}%
\usepackage{textcomp}%
\usepackage{manyfoot}%
\usepackage{booktabs}%
\usepackage{algorithm}%
\usepackage{algorithmicx}%
\usepackage{algpseudocode}%
\usepackage{listings}%

\theoremstyle{thmstyleone}%
\theoremstyle{thmstyletwo}%

\theoremstyle{thmstylethree}%

\begin{document}

\title[TokaGrad: End-to-end differentiable tokamak simulator for L-to-H full scenario modeling]{TokaGrad: End-to-end differentiable tokamak simulator for L-to-H full scenario optimization}

\author*[1]{\fnm{Jaemin} \sur{Seo}}\email{jseo@cau.ac.kr}

\affil*[1]{\orgdiv{Department of Physics}, \orgname{Chung-Ang University}, \orgaddress{\city{Seoul}, \postcode{06974}, \country{Republic of Korea}}}

\abstract{
As fusion energy moves from theoretical feasibility toward commercialization, design of new reactor concepts, autonomous tokamak control, and high-performance scenario optimization are becoming increasingly important. Traditionally, such optimization tasks have relied on costly trial-and-error or brute-force parameter searches, based on black-box experiments or simulations. More recently, machine learning has enabled data-driven approaches for tokamak control and scenario development. In parallel, advances in differentiable programming are changing the paradigm of numerical simulation. Unlike conventional simulations, which are typically executed as locally connected step-by-step procedures, differentiable simulation represents the entire simulation pipeline as a connected computational graph. In such a framework, machine parameters, actuator waveforms, and plasma responses are linked through differentiable operations, allowing Jacobians to propagate across the full simulation. This enables direct gradient-based control and optimization using the internal sensitivities of the simulator, rather than treating the simulator as a black box. Here, we present TokaGrad, an end-to-end differentiable tokamak transport simulator for full-scenario modeling, including ramp-up, L-mode operation, and H-mode access. TokaGrad self-consistently integrates differentiable models for plasma equilibrium, transport, heating, L-H transition, and pedestal formation. To our knowledge, this is the first differentiable tokamak simulator capable of self-consistently modeling dynamic full-discharge scenarios where actuators and plasma evolve together with equilibrium, pedestal, and confinement-regime transitions. We demonstrate that, when coupled to gradient-based optimizers, TokaGrad enables reactor-design optimization, actuator control, and full-scenario waveform optimization. This framework provides a pathway toward automated, differentiable optimization of burning-plasma scenarios and reactor concepts.
}


\maketitle

\section{Introduction}\label{sec1}

Fusion energy aims to reproduce on Earth the physical process that powers the Sun, offering a potential solution to two of the most pressing challenges of the coming decades, the rapidly increasing demand for energy and the need to reduce carbon emissions. As magnetic fusion research moves from the demonstration of scientific feasibility toward reactor-scale operation and commercialization, the ability to design new reactor concepts and optimize high-performance plasma scenarios is becoming increasingly important. This transition is reflected in the growing interest in DEMO-class reactors \cite{KWON2020_kdemo}, compact high-field tokamaks \cite{Rodriguez-Fernandez_2022_sparc}, and spherical tokamaks \cite{McNamara_2024_st40}. In parallel, advanced operating scenarios are being developed to maximize fusion performance within a given machine design, such as FIRE mode \cite{Han2022_fire, Na_2026_fire} and Super H-mode \cite{Snyder_2019_superH}. However, the design and optimization of such devices and scenarios remain challenging because reactor-grade tokamak operation involves many coupled control variables, nonlinear plasma responses, and simultaneous physics and engineering constraints.

Historically, tokamak scenario development has relied heavily on expensive experimental trial and error, expert-driven waveform tuning, and low-dimensional parameter scans. Integrated modeling tools such as ASTRA \cite{Tardini_2026_astra8}, TRANSP \cite{PANKIN2025_transp}, TRIASSIC \cite{Lee_2021_triassic}, and related transport simulation frameworks have greatly improved the ability to perform virtual experiments and interpret plasma behavior. Nevertheless, even with these tools, scenario design often remains a human-in-the-loop process in which actuator waveforms, boundary conditions, plasma configurations, and model assumptions are manually adjusted until an acceptable solution is obtained. This workflow becomes increasingly inefficient as the number of actuators, objectives, and constraints grows, and in reactor-grade devices, brute-force search or manual tuning becomes impractical.

Recent advances in machine learning have introduced new optimization strategies into plasma control and scenario development. Bayesian optimization, reinforcement learning, and data-driven modeling have been applied to actuator control \cite{Degrave2022_nature, Seo2024_nature}, disruption avoidance \cite{Yang_2025_disruption, Wang2025_disruption}, performance prediction \cite{NAM2025_ml_prediction, Yang2026_ml_prediction}, profile control \cite{Abbate_2023_control, Abbate_2025_control, Rothstein_2026_control}, and reactor design \cite{jskim2024_rl_reactor, bkim2026_bayesian_reactor}. These approaches provide powerful alternatives to manual tuning, especially when the search space is high-dimensional or the plasma response is strongly nonlinear. However, many of these methods still treat the underlying simulator or experiment as a ``black box.'' They typically require repeated forward evaluations and infer sensitivities indirectly through sampling, exploration, or learned policies. While effective in many settings, black-box optimization can become computationally expensive when applied to full-discharge, multi-actuator, multi-objective scenario optimization.

In parallel with these developments, differentiable programming and automatic differentiation (AD) have begun to transform the role of numerical simulation \cite{NEURIPS2020_jax_differentiable_sim, Newbury_2024_differentiable_sim}. In a differentiable simulation framework, the full numerical pipeline from initial condition to final state is represented as a connected differentiable computational graph. Inputs, intermediate states, physics closures, actuator waveforms, machine parameters, and performance metrics are connected by operations through which derivatives can propagate by the chain rule, in a manner analogous to the flow of gradients through the layers of a neural network. This allows one to compute sensitivities such as the dependence of fusion gain on plasma-current waveform, the dependence of pedestal formation on heating trajectory, or the dependence of final stored energy on geometry parameters. More importantly, these gradients can be used directly for gradient-based control and optimization, including modern optimizers such as Adam \cite{kingma2017_adam}, without requiring a separate finite-difference scan for each control parameter. This capability is particularly attractive for tokamak control and scenario optimization. In a conventional finite-difference approach, estimating the gradient of an objective with respect to a high-dimensional actuator waveform requires perturbing many control variables independently and rerunning the simulation many times. The cost therefore increases rapidly with the number of control degrees of freedom. By contrast, AD can provide gradients through the full simulation graph with a computational cost that is much less sensitive to the number of control variables. This makes differentiable simulation a promising approach for optimizing full-discharge scenarios in which plasma current, heating power, and shaping vary continuously in time.

Motivated by this paradigm shift, differentiable tokamak simulation frameworks have recently begun to emerge. A notable example is TORAX \cite{citrin2024_torax}, a JAX-based \cite{NEURIPS2020_jax_differentiable_sim, deepmind2020_jax_optax} differentiable transport simulator designed for pulse simulation and optimization using coupled transport equations. TORAX represents an important step toward differentiable integrated modeling for tokamak plasmas. However, existing differentiable transport solvers have primarily focused on core plasma transport and prescribed scenario evolution. Extensions to reactor-relevant full-discharge modeling remain limited, especially for scenarios involving L-to-H transition, edge pedestal formation, and the self-consistent coupling of evolving plasma shape with equilibrium reconstruction. These features are essential for predicting and optimizing burning-plasma scenarios, where the access to H-mode, the pedestal structure, bootstrap current, alpha heating, and current-profile evolution are tightly coupled.

In this work, we introduce TokaGrad (a differentiable tokamak simulator), an end-to-end differentiable tokamak transport simulator for L-to-H mode full-scenario prediction and optimization. TokaGrad is designed to simulate full-discharge and dynamic tokamak scenarios while preserving the gradients among the machine, actuator, and plasma parameters. The framework integrates differentiable models for equilibrium geometry, transport fluxes, heat and particle sources, and pedestal formation. By connecting these modules within a JAX-based differentiable computational graph, TokaGrad enables not only forward simulation of a pulse scenario, but also automatic-differentiation-based optimization of actuator controls, scenario trajectories, and reactor-design parameters.

The central objective of this paper is to demonstrate that differentiable integrated modeling can provide a practical route toward automated tokamak scenario design. In Section \ref{sec2}, we first describe the differentiable transport framework and the physics modules implemented in TokaGrad. We then present interactive simulation capabilities and benchmark the framework against existing transport simulation results in Section \ref{sec3}. We also demonstrate full-discharge simulation of an ITER-relevant L-to-H scenario. Finally, in Section \ref{sec4}, we show how AD enables gradient-based actuator control, full-scenario waveform optimization, and reactor-design optimization. These demonstrations suggest that TokaGrad can serve as a differentiable platform for predictive scenario design and optimization in future burning-plasma devices.

\section{Differentiable integrated modeling}\label{sec2}

The development of differentiable programming has opened new possibilities for using computational models not only as forward solvers, but also as gradient-carrying components in optimization and inverse-design workflows. In particular, differentiable physics has emerged as a framework in which physical laws, governing equations, and numerical solvers are embedded into differentiable computational graphs \cite{NEURIPS2020_jax_differentiable_sim, Newbury_2024_differentiable_sim}. This approach allows the outputs of a physical model to be differentiated with respect to inputs, parameters, boundary conditions, and control variables, thereby enabling sensitivity analysis, parameter inference, and gradient-based optimization within a unified computational framework.

A widely studied class of differentiable physics methods uses neural networks (NNs) as the primary computational backbone. Physics-informed neural networks \cite{RAISSI2019_pinn} and neural operators \cite{Azizzadenesheli2024_neural_operator} encode physical constraints into the training objective or architecture of a differentiable model. These methods have been applied across many areas of science and engineering, and have shown particular promise for inverse problems such as estimating hidden coefficients and reconstructing internal states from sparse observations \cite{Seo2024_pinn_optim, seo2024_pre_pinn, SEO2024_net_pinn}. In such approaches, the governing equations provide physical regularization, while the neural network supplies a flexible differentiable representation of the solution or operator.

However, despite their flexibility, NN-based physics models also inherit several challenges from data-driven deep learning. They generally exhibit poor extrapolation capability, high sensitivity to random seed and hyperparameters, and low reproducibility even with the same dataset. An alternative route is differentiable simulation, in which the backbone is not an NN, but a conventional numerical solver implemented as a differentiable computational graph. In this case, finite-difference, finite-element, or matrix-solver operations are constructed from differentiable primitives, allowing derivatives to propagate through the numerical time evolution. The governing equations remain explicit in the solver structure, while AD provides the Jacobians connecting inputs, intermediate states, and output metrics. It is important to distinguish this use of automatic differentiation from the finite differences used to discretize differential equations. The finite-difference or finite-volume scheme defines the numerical approximation to the physical equations, whereas AD computes derivatives of the resulting computational graph with respect to model parameters or control variables.

\begin{figure}[t!]
\centering
\includegraphics[width=3.5in]{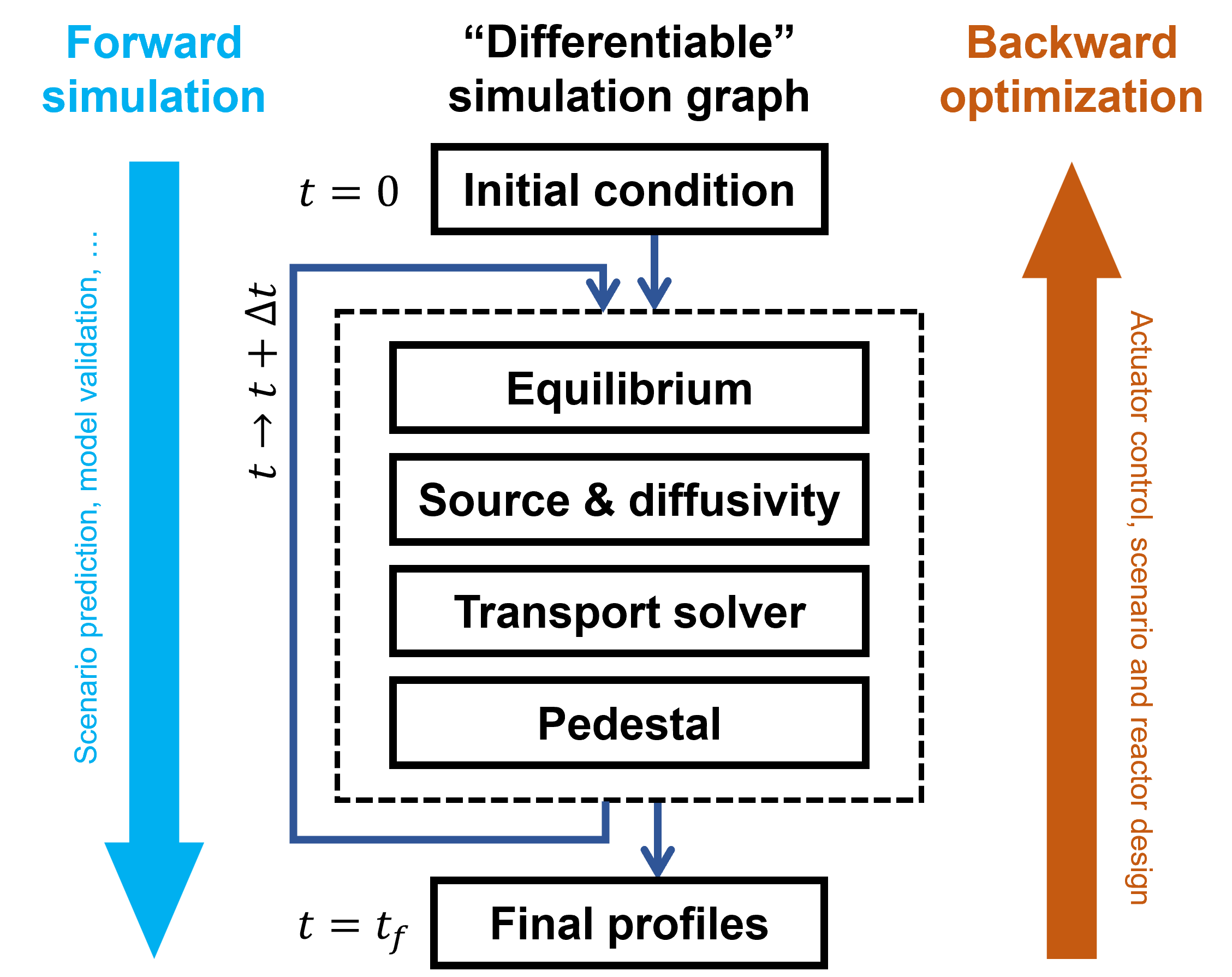}
\caption{Differentiable tokamak simulation framework. This framework, composed of a differentiable graph, allows not only typical forward transport simulations but also backward optimizations of actuator control, scenario waveform, and reactor design.}\label{fig1}
\end{figure}

TokaGrad follows this differentiable-simulation approach. Rather than replacing tokamak transport modeling with a purely data-driven NN, we implement the conventional structure of an integrated tokamak transport simulator in a JAX-based differentiable framework. The core of the model is a set of coupled one-dimensional transport equations for plasma profiles, evolved in a flux-coordinate representation appropriate for axisymmetric tokamak geometry. Around this transport backbone, we integrate differentiable or differentiability-preserving models for equilibrium geometry, turbulent and neoclassical transport, heat and particle sources, and pedestal formation, as summarized in Table \ref{tab:models}. These modules are coupled through the evolving plasma state, actuator waveforms, and machine parameters, forming a connected computational graph from scenario inputs to final plasma performance. Integration in a differentiable framework enables not only forward simulation but also backward optimization based on the gradient between each node, as shown in Figure \ref{fig1}.

\begin{table}[h!]
\renewcommand{\arraystretch}{1.3} 
\centering
\caption{Differentiable modules integrated into TokaGrad}
\label{tab:models}
\begin{tabular}{lccc}
\hline
  & Available models & Gives \\
\hline
Main body & TokaGrad & $T_{\rm e,i}$, $n_{\rm e}$, $\psi$, $\textbf{j}_{\rm ind}$ \\
Equilibrium & EQDSK \cite{LUTJENS1996_chease}, Miller/moment solver \cite{Miller_1998} & $\Phi_{\rm b}$, $V'$, $g_{0...3}$, $F$, $q$ \\
Turbulent transport &  TGLF-NN \cite{Meneghini_2017_tglfnn_eped1nn}, QLKNN \cite{qlknn_2020}, BgB \cite{Erba_1997_bgb} & $D_{\rm e}$, $\chi_{\rm e,i}$, $V_{\rm e}$, $q_{\rm conv}$ \\
Neoclassical physics & NEO-NN \cite{Meneghini_2017_tglfnn_eped1nn}, Sauter-Angioni \cite{sauter_1999_neoclassical, angioni_2000_neoclassical} & $D_{\rm e}$, $\chi_{\rm e,i}$, $\sigma_\parallel$, $\textbf{j}_{\rm BS}$ \\
L-H transition & Martin \cite{Martin_2008_threshold}, Delabie \cite{Delabie_2026_threshold} & L- or H-mode regime \\
Pedestal & EPED1-NN \cite{Meneghini_2017_tglfnn_eped1nn}, alpha-critical \cite{Onjun_2002_alpha_crit} & $T_{\rm ped}$, $n_{\rm ped}$ (if H-mode) \\
Heating \& current & Ohmic, alpha, e-i exchange, & $P_{\rm e,i}$, $\textbf{j}_{\rm CD}$\\
 & radiation, gaussian auxiliary & \\
\hline
\end{tabular}
\end{table}

\subsection{Differentiable transport framework}\label{sec2.1}

The central advantage of a differentiable transport framework is that the inputs and outputs of the transport equations are connected within a single computational graph. Geometry, actuator waveforms, heating sources, boundary conditions, and model parameters all enter the evolution equations as differentiable inputs, while plasma profiles and performance metrics are obtained as differentiable outputs. AD can therefore be used to compute the Jacobians between these quantities without externally perturbing each parameter. For example, sensitivities such as ${\partial P_{\rm fus}(t_{\rm f})}/{\partial I_{\rm p}(t)}$ and ${\partial Q(t_{\rm f})}/{\partial P_{\rm aux}(t)}$ can be evaluated directly from the simulation graph, where $P_{\rm fus}(t_{\rm f})$ and $Q(t_{\rm f})$ are the final fusion power and gain, $P_{\rm aux}(t)$ is an auxiliary-heating waveform, and $I_{\rm p}(t)$ is the plasma-current waveform. These sensitivities can then be used either for interpretation, by identifying which actuators or design parameters most strongly affect the final plasma state, or for optimization, by updating the controls in the direction that improves a chosen objective.

The emergence of neural-network surrogate models further strengthens this approach. Many high-fidelity physics models used in integrated modeling, such as TGLF \cite{staebler_2007_tglf1, kinsey_2008_tglf2} and QuaLiKiz \cite{Citrin_2017_qualikiz} for turbulent transport, NEO \cite{Belli_2015_NEO} for neoclassical transport, and EPED1 \cite{Snyder_2009_eped1} for pedestal, are computationally expensive and are not designed to be differentiated through. They include iterative solvers, conditional branching, or non-smooth operations that break the differentiable path required for gradient-based optimization. Recently, NN surrogates such as TGLF-NN, QLKNN, NEO-NN, and EPED1-NN have been developed to accelerate these physics models \cite{Meneghini_2017_tglfnn_eped1nn, qlknn_2020}. Although these surrogates were originally motivated by speed, they also provide differentiable mappings from plasma parameters to transport, neoclassical, or pedestal quantities. When these surrogate models are converted to a JAX-compatible form and embedded into a differentiable transport solver, the gradients can propagate not only through the numerical transport equations, but also through the physics closures themselves. Thus, the present availability of NN surrogates makes it possible to construct an integrated tokamak simulator that is end-to-end differentiable from actuator and design inputs to final plasma performance.

The core of TokaGrad is a set of coupled one-dimensional transport equations written in a normalized radial flux coordinate $\rho \equiv \rho_{\rm tor} / \rho_{\rm tor,b}$. Here, $\rho_{\rm tor} = \sqrt{\Phi / \pi B_T}$ is the effective minor radius and $\rho_{\rm tor,b}$ is its boundary value, where $\Phi$ is the toroidal magnetic flux and $B_T$ is the toroidal field. The evolved plasma state consists of electron and ion temperature $T_{\rm e, i}$, electron density $n_{\rm e}$, and a poloidal magnetic flux $\psi$. The radial coordinate is associated with flux-surface geometry through the differential volume element $V'=\partial V / \partial \rho$ and metric factors such as $g_{0,1}$, which represent the geometric weights multiplying radial fluxes. In this representation, the particle transport equation is written in conservative form as

\begin{equation}
    \left(\frac{\partial}{\partial t} 
    - 
    \frac{\dot{\Phi}_{\rm b}}{2 \Phi_{\rm b}} \frac{\partial}{\partial \rho} \rho\right)
    \left[V' n_{\rm e}\right] 
    =
    \frac{\partial}{\partial \rho}
    \left[
    D_{\rm e}\frac{g_1}{V'}\frac{\partial n_{\rm e}}{\partial \rho}
    -
    g_0 V_{\rm e} n_{\rm e}
    \right]
    +
    V' S_n ,
\label{eq_density_diffusion} \\
\end{equation}

where $D_{\rm e}$ is the particle diffusivity, $V_{\rm e}$ is a convective velocity, and $S_n$ is the particle source. $\Phi_{\rm b}$ denotes the boundary toroidal flux. The term proportional to $\dot{\Phi}_{\rm b}/\Phi_{\rm b}$ represents the effect of time-varying geometry or flux-surface volume during dynamic scenarios such as current ramp-up and plasma-shape evolution. The electron and ion temperature equations are solved in a conservative energy form. For species $s\in{{\rm e},{\rm i}}$, TokaGrad evolves the variable $V'^{5/3} n_s T_s$, which is convenient for preserving the geometric compression and expansion terms associated with evolving flux-surface volume. The implemented heat equation can be expressed schematically as

\begin{equation}
\frac{3}{2} V'^{-5/3}
\left(
\frac{\partial}{\partial t}
-
\frac{\dot{\Phi}_{\rm b}}{2\Phi_{\rm b}}
\frac{\partial}{\partial \rho}\rho
\right)
\left[
V'^{5/3} n_s T_s
\right]
=
\frac{1}{V'}
\frac{\partial}{\partial \rho}
\left[
\chi_s n_s \frac{g_1}{V'}
\frac{\partial T_s}{\partial \rho}
-
g_0 q_{{\rm conv},s}T_s
\right]
+
P_s .
\label{eq_temperature_diffusion} \\
\end{equation}

Here $\chi_s$ is the thermal diffusivity, $q_{{\rm conv},s}$ is a heat-convection coefficient, $P_s$ is the net temperature source rate including external heating, collisional electron-ion exchange, alpha heating, and radiation losses.

The magnetic-flux evolution is treated through a current-diffusion model coupled to the evolving conductivity, bootstrap current, and external current-drive terms. In a generic form, the inductive flux equation can be written as

\begin{equation}
\frac{16\pi^2\mu_0\sigma_\parallel\rho\Phi_{\rm b}^2}{F^2}
\left(\frac{\partial \psi}{\partial t}-
\frac{\rho\dot{\Phi}_{\rm b}}{2\Phi_{\rm b}}
\frac{\partial \psi}{\partial \rho}\right)
=
\frac{\partial}{\partial \rho}
\left(
\frac{g_2 g_3}{\rho} \frac{\partial \psi}{\partial \rho}
\right)
-
\frac{8\pi^2V'\mu_0\Phi_{\rm b}}{F^2}
\langle{\textbf{B}\cdot (\textbf{j}_{\rm BS}+\textbf{j}_{\rm CD})}\rangle ,
\label{eq_psi_diffusion} \\
\end{equation}

where $\sigma_\parallel$ is the parallel conductivity and $F=RB_T$ is the poloidal current function, where $R$ is the major radius and $B_T$ is the toroidal field. $g_{2,3}$ denote the relevant geometric factors and $\textbf{j}_{\rm NI}=\textbf{j}_{\rm BS}+\textbf{j}_{\rm CD}$ is non-inductive current sources. The total current density is then reconstructed from the inductive ($\textbf{j}_{\rm ind}$), bootstrap ($\textbf{j}_{\rm BS}$), and auxiliary current components ($\textbf{j}_{\rm CD}$), and the safety-factor profile $q$ is updated consistently with the evolving equilibrium geometry. This coupling is important because the current profile affects magnetic shear, transport closures, pedestal stability, and the subsequent plasma response.

The geometric factors $g_{0,1,2,3}$ \cite{citrin2024_torax} are defined as

\begin{equation}
    g_0=\langle\nabla V\rangle, \quad g_1=\langle (\nabla V)^2\rangle, \quad g_2=\langle(\nabla V)^2/R^2 \rangle, \quad g_3=\langle 1/R^2\rangle.
\label{eq_geometric_factors} \\
\end{equation}

Numerically, TokaGrad advances these coupled transport equations using a semi-implicit finite-volume method by default, while explicit, full-implicit, predictor-corrector, and Newton-Raphson schemes are also available. In the semi-implicit method, transport coefficients, geometry, and pedestal are evaluated from the current or reference plasma state, while the diffusive part of the transport equations is advanced implicitly. For a generic profile variable $y$, the semi-implicit update after $\Delta t$ can be written as

\begin{equation}
    \left(I-\Delta t \cdot L[y^{k}]\right)y^{k+1}
    =
    y^{k}+\Delta t \cdot S[y^{k}],
\label{eq_implicit} \\
\end{equation}

where $L[y^k]$ is the finite-volume diffusion operator evaluated using the transport coefficients and geometry from the reference state, and $S[y^k]$ denotes explicit sources. This update is more stable than a fully explicit method because the stiff radial diffusion operator is treated implicitly, while avoiding the cost and complexity of solving a fully nonlinear implicit system at every step. When stronger nonlinear convergence is required, the same semi-implicit update can be repeated in a Picard iteration, using the updated state as the reference state for the next iteration.

The finite-volume discretization leads to tridiagonal radial operators for the one-dimensional diffusion equations. TokaGrad therefore solves the implicit step using a differentiable tridiagonal matrix solver based on the Thomas algorithm, rather than constructing and factorizing dense matrices. This gives an $O(N_\rho)$ solve for each radial profile, where $N_\rho$ is the number of radial grid points, and preserves compatibility with JAX transformations such as just-in-time compilation, vectorization, and automatic differentiation. The time-integration loop is implemented using JAX-compatible control flow, so that the full pulse evolution can remain inside the differentiable computation graph.

Several numerical treatments are introduced to improve robustness while maintaining differentiability. Boundary conditions at the magnetic axis and plasma edge are imposed directly in the finite-volume operators, with no-flux symmetry at the axis and prescribed edge temperature or density values at the last closed flux surface. Source terms can be evaluated with semi-implicit local sink treatment to avoid unphysical overshoots when strong radiation or heat exchange is present. For differentiable optimization, non-smooth operations such as hard switching, clipping, or pedestal enforcement can be replaced by smooth approximations, so that gradients remain informative during automatic differentiation. Conversely, for pure forward simulation, stricter limiters can be used to improve numerical safety.

This transport framework provides the backbone to which the remaining physics modules are coupled. The transport equations cannot be closed by themselves: they require geometry and metric factors from an equilibrium model, transport coefficients from turbulent and neoclassical closures, heat and particle sources from heating and fueling models, and pedestal or L-H transition models to describe confinement-regime changes. These components are described in the following subsections.

\subsection{Geometry}\label{sec2.2}

Solving Equations \ref{eq_density_diffusion}--\ref{eq_psi_diffusion} requires geometric information ($\Phi_{\rm b}$, $V'$, $g_{0...3}$, and $F$) that connects the one-dimensional radial coordinate used in the transport solver to the two-dimensional poloidal cross-section of the tokamak plasma. Because a tokamak is approximately axisymmetric, the toroidal angle can be eliminated from the transport problem. The remaining two-dimensional poloidal structure can then be reduced to a one-dimensional radial problem by using magnetic flux surfaces as the radial coordinate. This is the standard basis of a 1.5D transport model: the plasma profiles are evolved only along a normalized flux coordinate $\rho$, while the effects of the poloidal geometry enter through flux-surface-averaged metric factors.

TokaGrad provides two complementary approaches for modeling the two-dimensional equilibrium geometry. The first approach uses externally generated equilibria in the standard EQDSK format \cite{LUTJENS1996_chease}. EQDSK files are commonly produced by Grad-Shafranov equilibrium reconstruction solvers such as EFIT \cite{Lao_2022_efit} and CHEASE \cite{LUTJENS1996_chease}. They contain the two-dimensional poloidal-flux map $\psi(R,Z)$, pressure and current-related profiles, and other equilibrium quantities. From this information, TokaGrad extracts flux-surface moments and metric quantities such as surface volume, differential volume, effective minor radius, elongation, triangularity, magnetic-axis shift, and radial metric factors.

The EQDSK-based geometry path is useful when realistic equilibria are available from experiment or an external Grad-Shafranov solver. It can represent shaped plasmas, X-point configurations, and realistic flux-surface structures more accurately than a reduced analytic model. It is therefore suitable for forward modeling and benchmarking against existing integrated modeling workflows. However, because the equilibrium is prescribed by an external file or sequence of files, the geometry itself is not generally solved self-consistently inside the differentiable graph. In this mode, differentiability is preserved through the transport evolution for a given geometry, while gradients with respect to arbitrary geometry changes are limited unless the geometry sequence is parameterized in a differentiable way.

The second approach is a reduced fixed-boundary geometry model based on Miller-like flux-surface parameterization and moment reconstruction \cite{Miller_1998}. In this model, each flux surface is described by a small set of shape parameters, such as geometric major radius $R_0$, minor radius $a$, elongation $\kappa$, triangularity $\delta$, and Shafranov shift $\Delta$. A representative flux surface can be written as

\begin{equation}
R(\rho,\theta)
=
R_0 + \Delta(\rho)
+
a\rho \cos\left[\theta+\delta(\rho)\sin\theta\right], \quad
Z(\rho,\theta)
=
\kappa(\rho)a\rho \sin\theta ,
\label{eq_miller} \\
\end{equation}

where $\theta$ is the poloidal angle. The flux-surface geometry is then obtained by evaluating moments and metric factors directly from this parameterized surface. Unlike a full Grad-Shafranov solve, this procedure does not require an iterative equilibrium calculation. It is therefore much faster and more naturally compatible with automatic differentiation. The reduced geometry model is not intended to replace a free-boundary equilibrium solver. It does not compute coil responses, scrape-off-layer geometry, or the full force-balance solution of the Grad-Shafranov equation. Instead, it provides a differentiable representation of the dominant geometric effects needed by the transport equations. To incorporate part of the equilibrium response in a reduced form, the magnetic-axis position is allowed to shift according to an analytic Shafranov-shift model. In TokaGrad, this is represented schematically as

\begin{equation}
\Delta(\rho)
=
f_\Delta(\rho) \epsilon a
\left(\beta_{\rm p}+\frac{l_{\rm i}}{2}-\frac{1}{2}\right),
\label{eq_shift} \\
\end{equation}

where $\epsilon=a/R_0$ is the inverse aspect ratio, $\beta_{\rm p}$ is the poloidal beta, $l_{\rm i}$ is the internal inductance, and $f_\Delta(\rho)$ is a reduced-model coefficient. This form is not a substitute for a Grad-Shafranov equilibrium calculation, but it captures the leading tendency of high-pressure plasmas to shift the magnetic axis outward and allows the geometry to respond smoothly to changes in plasma pressure and current-profile quantities.

The fixed-boundary moment geometry has two advantages for the purposes of differentiable scenario optimization. First, it is computationally inexpensive because all geometric factors are obtained from analytic or quadrature-based operations rather than an iterative equilibrium solve. Second, the shape parameters can be treated as differentiable inputs. This makes it possible to compute gradients of plasma performance with respect to machine-design or shape parameters, such as $R_0$, $a$, and $\kappa$. Such gradients are essential for design-space exploration and reactor-shape optimization.

The two geometry modes therefore serve different purposes. The EQDSK-based mode is more realistic and is appropriate when externally reconstructed or designed equilibria are available. The fixed-boundary moment mode is less complete as an equilibrium model, but is faster, smoother, and more suitable for AD-based shape and scenario optimization. TokaGrad supports both modes so that realistic prescribed-equilibrium simulations and differentiable geometry-optimization studies can be performed within the same transport framework.

\subsection{Transport closure model}\label{sec2.3}

Particle and heat transport in tokamak plasmas are governed primarily by turbulent and neoclassical processes. The transport equations in Section \ref{sec2.1} therefore require closure models that provide the particle and thermal diffusivity, convective fluxes, and non-inductive current components as functions of the evolving plasma state. In TokaGrad, the transport closure is evaluated at each time step from local profile quantities such as $T_{\rm e,i}$ and $n_{\rm e}$. The resulting transport coefficients are then passed to the semi-implicit transport solver.

For turbulent transport, TokaGrad provides three classes of closure models. The first is an empirical Bohm-gyroBohm (BgB) model. Because this model is composed of explicit arithmetic operations, it is naturally fast and differentiable. However, the BgB model cannot accurately represent the broad range of turbulent transport regimes encountered in reactor-relevant plasmas, including ion-temperature-gradient and trapped-electron-mode turbulence.

For higher-fidelity turbulent transport, gyrokinetic or gyrofluid models such as QuaLiKiz \cite{Citrin_2017_qualikiz} and TGLF \cite{staebler_2007_tglf1} are commonly used in integrated modeling. These models can capture important microinstability-driven transport channels and provide more physics-based estimates of heat and particle fluxes. However, directly coupling such models to a time-dependent integrated simulator is computationally expensive because the transport coefficients must be updated repeatedly over all radial grid points and time steps. In addition, the original implementations are not generally suitable for end-to-end automatic differentiation due to their non-differentiable internal algorithms. To overcome these limitations, TokaGrad uses NN surrogate transport models, including QLKNN \cite{qlknn_2020} and TGLF-NN \cite{Meneghini_2017_tglfnn_eped1nn}. These models approximate the input-output maps of the corresponding gyrokinetic or gyrofluid transport models while reducing the evaluation cost by orders of magnitude. From the perspective of differentiable simulation, their most important feature is that they are themselves differentiable computational graphs. Once converted to a JAX-compatible implementation, the surrogate model can be evaluated inside the same graph as the transport solver, allowing gradients to propagate from final plasma performance metrics through the turbulent transport closure back to input parameters. This makes NN transport surrogates particularly well matched to the goal of end-to-end differentiable integrated modeling.

For neoclassical transport, TokaGrad supports both analytic neoclassical formulas and NN surrogate closures. In the analytic path, Sauter-Angioni formulas are used to estimate neoclassical diffusivities and the bootstrap current \cite{sauter_1999_neoclassical, angioni_2000_neoclassical}. In the surrogate path, NEO-NN provides an NN approximation to the neoclassical transport model NEO. As with the turbulent transport surrogates, a JAX-compatible NEO-NN implementation allows the neoclassical closure to remain inside the differentiable simulation graph.

By combining empirical, surrogate-based, and analytic closure options, TokaGrad can be used across different levels of fidelity and computational cost. Empirical models provide fast and robust differentiable baselines, neural-network transport surrogates provide higher-fidelity turbulent and neoclassical closures while preserving differentiability, and analytic neoclassical formulas provide interpretable bootstrap and collisional transport estimates. This modular structure allows the same differentiable transport framework to support rapid testing, benchmark studies, full-scenario prediction, and gradient-based optimization.

\subsection{Heat and particle sources}\label{sec2.4}

In addition to transport losses, tokamak plasma evolution is strongly determined by heat and particle sources. Turbulent and neoclassical transport continuously remove energy and particles from the confined plasma, and these losses must be balanced by external heating, self-heating, fueling, and current-drive mechanisms in order to reach and sustain reactor-relevant conditions. In TokaGrad, heat and particle sources are implemented as differentiable or differentiability-preserving modules coupled to the transport equations described in Section \ref{sec2.1}.

The electron and ion heat equations include several source and sink terms. Ohmic heating is computed from the evolving current density and plasma resistivity, and is primarily deposited into the electron channel. Collisional electron-ion heat exchange transfers energy between electrons and ions, tending to equilibrate $T_{\rm e}$ and $T_{\rm i}$ on the collisional energy-exchange time scale. In deuterium-tritium plasmas, fusion alpha heating is also included. The fusion reaction rate is evaluated using the Bosch-Hale fitting formula \cite{Bosch_1992_fusion_reactivity}, and the resulting alpha-particle power is distributed between electron and ion channels using a slowing-down partition theory. This coupling is essential for burning-plasma scenarios, where auxiliary heating can trigger increased fusion power, which then produces alpha heating and modifies the subsequent confinement-regime evolution.

Auxiliary heating is currently modeled using prescribed radial deposition profiles. For a heating source $k$, the deposited power density is written as a normalized Gaussian profile,

\begin{equation}
P_k(\rho,t)
=
P_{k,{\rm tot}}(t)
\frac{
\exp\left[-(\rho-\rho_k(t))^2/2\sigma_k^2(t)\right]
}{
\int_0^1
\exp\left[-(\rho-\rho_k(t))^2/2\sigma_k^2(t)\right]
V'(\rho,t),d\rho
},
\label{eq_auxiliary} \\
\end{equation}

where $P_{k,{\rm tot}}(t)$ is the total injected power, $\rho_k(t)$ is the deposition location, and $\sigma_k(t)$ is the deposition width. The profile is normalized so that its volume integral equals the externally specified input power. This reduced treatment is less detailed than external heating modules such as NUBEAM \cite{PANKIN2004_nubeam} for neutral beam injection. However, it is computationally inexpensive, smooth, and well suited to a differentiable optimization framework. In future work, these reduced profiles can be replaced or augmented by differentiable neural-network surrogate heating models \cite{Boyer_2019_nbi_NN1, MOROSOHK2021_nbi_nn2, WANG2023_nbi_nn3, Rothstein_2024_nbi_nn4}.

Radiative power losses are also included as sink terms in the electron heat equation. TokaGrad implements bremsstrahlung, synchrotron, and line-radiation losses \cite{huba1998nrl, Albajar_2001_synchrotron}. These terms depend on local plasma quantities such as density, temperature, magnetic field, effective charge, impurity content, and geometric parameters. The net electron and ion source terms can therefore be written schematically as

\begin{equation}
P_{\rm e}
=
P_{\rm ohm}
+
P_{{\rm aux},e}
+
P_{\alpha,e}
-
P_{{\rm rad}}
-
P_{{\rm ei}},
\qquad
P_{\rm i}
=
P_{{\rm aux},i}
+
P_{\alpha,i}
+
P_{{\rm ei}},
\label{eq_heating} \\
\end{equation}

where $P_{\rm ohm}$ is Ohmic heating, $P_{{\rm aux},s}$ is auxiliary heating deposited to species $s$, $P_{\alpha,s}$ is alpha heating, $P_{\rm rad}$ is the total radiative loss, and $P_{\rm ei}$ denotes the electron-ion heat exchange term.

The treatment of particle sources is more subtle than that of heat sources. In experiments, the particle inventory is affected by gas puffing, pellet injection, wall retention, recycling, pumping, neutral penetration, and edge transport. These processes have substantial uncertainties and are often not controlled by specifying a known particle source profile directly. Instead, experiments commonly use density feedback control \cite{Mlynek_2011_density_feedback_control, Zheng_2013_density_feedback_control, JANKY2015_density_feedback_control, JUHN2020_density_feedback_control}, in which fueling actuators are adjusted to track a target line-averaged density or Greenwald fraction. TokaGrad follows this scenario-level viewpoint by providing density-control options based on target density quantities rather than requiring a precisely prescribed fueling source.

One option is to evolve the density profile using a particle transport equation with an edge-localized feedback source. In this mode, the code computes the deviation between the evolving density and a target density measure, such as the Greenwald fraction $f_{\rm GW}$, and adjusts the fueling source to reduce this error. The target density is related to the Greenwald density, $n_{\rm GW}={I_{\rm p}}/{\pi a^2}$, where $I_{\rm p}$ is the plasma current in MA, $a$ is the minor radius in m, and $n_{\rm GW}$ is expressed in units of $10^{20}$ ${\rm m}^{-3}$. The target line-averaged or volume-averaged density is then set by $f_{\rm GW} n_{\rm GW}$. This allows the density to evolve consistently with the changing plasma current and size during ramp-up and flattop phases.

A second option is a prescribed-shape density model. In this mode, the radial shape of the density profile is specified by an analytic or reference profile, and the entire profile is rescaled at each time step to match the target Greenwald fraction or line-averaged density. This approach is motivated by the relative robustness of density peaking expected in ITER and reactor-relevant scenarios, and by the fact that density control is often better represented at the scenario level than through a first-principles fueling model. Although this option does not predict detailed fueling, recycling, or pumping dynamics, it provides a stable and differentiable density evolution suitable for full-discharge optimization studies.

Together, these heat and particle source models provide the energetic and fueling inputs needed to simulate dynamic tokamak scenarios. External heating raises the plasma temperature and can trigger L-H transition, alpha heating introduces burning-plasma feedback, radiation losses limit temperature and power balance, and density control determines the collisionality, fusion reactivity, Greenwald fraction, and transport response. Because these source terms are embedded in the same differentiable graph as the transport and geometry models, TokaGrad can compute how changes in heating waveform, deposition location, density target, or machine parameters affect the final plasma state and performance.

\subsection{L-H transition and pedestal}\label{sec2.5}

Access to H-mode is essential for achieving high fusion gain in ITER and reactor-grade tokamak plasmas. Compared with L-mode operation, H-mode provides improved energy confinement through the formation of a pedestal in the temperature and density profiles. This pedestal raises the boundary condition for the core plasma and can substantially increase the stored energy and fusion performance. In burning-plasma scenarios, the pedestal also affects alpha heating, bootstrap current, edge stability, and the power crossing the separatrix. Therefore, a full-scenario simulator intended for reactor optimization must be able to model not only core transport, but also the transition from L-mode to H-mode and the subsequent evolution of the pedestal.

Empirically, the L-to-H transition occurs when the power crossing the separatrix exceeds a threshold value \cite{Martin_2008_threshold, Delabie_2026_threshold}. For ITER baseline operation, the predicted L-H threshold power is expected to be comparable to the available auxiliary heating power, making H-mode access a marginal and scenario-dependent problem. This is important because a plasma that could sustain H-mode once the pedestal and alpha heating are established may still fail to access H-mode during ramp-up if the heating trajectory, density evolution, current waveform, or shape evolution is not favorable. Consequently, waveform optimization is not only a matter of maximizing flattop performance, but also of finding a dynamically accessible path through L-mode, transition, and H-mode phases under finite actuator limits.

Modeling the pedestal in integrated transport simulation is challenging. In many transport simulation studies, the pedestal top is prescribed or imposed as a boundary condition for the core transport equations, and the pedestal region itself is not evolved dynamically \cite{Lee_2021_triassic, Schramm_2025_pedestal_fixed, BRAY2026_pedestal_fixed}. This treatment can be adequate for interpreting fixed H-mode phases when pedestal quantities are known or prescribed. However, it is less suitable for predictive full-discharge simulation because the pedestal height and width depend sensitively on the evolving plasma state. In particular, pedestal fueling, edge transport, bootstrap current, and alpha-heating feedback can all influence whether the plasma accesses and sustains H-mode. For this reason, TokaGrad includes differentiable pedestal and L-H transition models as part of the integrated simulation graph, rather than treating the pedestal only as a fixed external boundary condition.

The L-H transition model compares the power crossing the separatrix, $P_{\rm sep}$, with an empirical threshold power, $P_{\rm LH}$. The threshold can be evaluated using empirical scalings such as the Martin scaling \cite{Martin_2008_threshold} or the Delabie scaling \cite{Delabie_2026_threshold},

\begin{equation}
P_{\rm LH}
=
\mathcal{P}_{\rm LH}
\left(
n_{\rm e}, B_{\rm T}, S, M,  \cdots
\right),
\label{eq_lh_threshold} \\
\end{equation}

where $S$ is the plasma surface area and $M$ is the ion mass number. The separatrix power is computed from the plasma power balance,

\begin{equation}
P_{\rm sep}
=
P_{\rm aux}
+
P_{\rm ohm}
+
P_{\alpha}
-
P_{\rm rad}
-
\frac{dW_{\rm th}}{dt},
\label{eq_psep}
\end{equation}

where $W_{\rm th}$ is the thermal stored energy. Depending on the application, simplified forms of this expression can be used, for example by neglecting the transient stored-energy term or by using the absorbed power to the plasma. A strict L-H switch based on $P_{\rm sep}>P_{\rm LH}$ would introduce a discontinuity into the simulation, which is undesirable for automatic differentiation and gradient-based optimization. TokaGrad therefore uses a smooth transition variable,

\begin{equation}
h_{\rm LH}
=
\sigma
\left[
\frac{P_{\rm sep}-P_{\rm LH}}{\Delta P_{\rm LH}}
\right],
\qquad
\sigma(x)=\frac{1}{1+\exp(-x)},
\end{equation}

where $h_{\rm LH}\sim0$ corresponds to L-mode-like confinement, $h_{\rm LH}\sim1$ corresponds to H-mode-like confinement, and $\Delta P_{\rm LH}$ controls the smoothness of the transition. This smooth gate allows the transport coefficients, pedestal targets, and confinement-regime-dependent source terms to vary continuously through the transition. It also provides meaningful gradients near the threshold, enabling optimization algorithms to adjust actuator waveforms toward H-mode access rather than encountering a non-differentiable binary event.

Once the plasma enters the H-mode regime, TokaGrad constructs a pedestal target using either an alpha-critical reduced model or an EPED1-NN surrogate model. In the alpha-critical model \cite{Onjun_2002_alpha_crit}, the pedestal width and height are related to an edge stability constraint. A schematic form of the model is

\begin{equation}
\Delta_{\rm ped}
=
C_w \beta_{\rm p,ped}^{0.5}
\qquad
p_{\rm ped}
\simeq
p_{\rm sep}
+
\left|\frac{dp}{d\rho}\right|_{\rm crit}
\Delta_{\rm ped},
\end{equation}

where $\Delta_{\rm ped}$ is the pedestal width, $\beta_{\rm p,ped}$ is a pedestal poloidal-beta measure, $C_w=0.076$ is a width coefficient by kinetic-ballooning-mode constraints \cite{Snyder_2009_eped1}, $p_{\rm sep}$ is the separatrix pressure, and $|dp/d\rho|_{\rm crit}$ is the critical pressure gradient determined from an empirical stability criterion \cite{Onjun_2002_alpha_crit}, 

\begin{equation}
\alpha_{\rm crit} \equiv \frac{2\mu_0 R_0 q^2}{a B_T^2} \left| \frac{dp}{d\rho} \right|_{\rm crit} =0.4s\left[ 1+\kappa_{95}^2 (1+5\delta_{95}^2) \right].
\end{equation}

This reduced model captures the idea that the pedestal height is constrained by edge pressure-gradient stability and that the pedestal width grows with pedestal beta.

For higher-fidelity pedestal closure, TokaGrad can use an EPED1-NN model. EPED1 combines constraints associated with peeling-ballooning stability and kinetic-ballooning-limited pedestal width to predict pedestal height and width \cite{Snyder_2009_eped1}. The neural-network surrogate provides a fast differentiable mapping \cite{Meneghini_2017_tglfnn_eped1nn},

\begin{equation}
\left(
p_{\rm ped}, \Delta_{\rm ped}
\right)
=
\mathcal{E}_{\rm EPED1-NN}
\left(
I_{\rm p}, B_{\rm T}, R_0, a, \kappa, \delta,
n_{\rm ped}, Z_{\rm eff}, M, \beta_{\rm N}
\right),
\end{equation}

where $n_{\rm ped}$ is the pedestal density, $Z_{\rm eff}$ is the effective charge number, and $\beta_{\rm N}$ is the normalized plasma beta. Because the surrogate is implemented in JAX, the predicted pedestal height and width remain differentiable with respect to actuator and geometry parameters. The pedestal target is incorporated into the evolving profiles through a smooth enforcement model. Rather than imposing a discontinuous pedestal boundary condition, TokaGrad blends the core transport solution with a pedestal-shaped profile near the edge. A generic form for variable $y \in T_{\rm e,i}, n_{\rm e}$ is

\begin{equation}
y^{\rm new}(\rho)
=
{\rm smoothmax}\left[
y^{\rm tr}(\rho), h_{\rm LH}y_{\rm ped}(\rho)
\right],
\end{equation}
\begin{equation}
y_{\rm ped}(\rho)=y_{\rm sep} + \frac{y_{\rm height}}{2} \left[1-\tanh\left(\frac{\rho-\rho_{\rm ped}}{0.5w}\right)\right],
\end{equation}

where $y^{\rm tr}(\rho)$ is the transport-evolved temperature or density profile, $y_{\rm ped}(\rho)$ is the pedestal target profile, $w$ is the pedestal width, and $\rho_{\rm ped}=1-0.5w$ is the location of the pedestal center. The multiplication by $h_{\rm LH}$ ensures that the pedestal is absent in L-mode and gradually appears as the separatrix power exceeds the L-H threshold. This smooth pedestal formation is important for differentiable optimization because the gradient of final performance with respect to heating or density trajectory must propagate through the transition and pedestal response.

This differentiable treatment of L-H transition and pedestal formation is one of the central features of TokaGrad. It allows the simulator to represent full-discharge scenarios in which a plasma evolves from ramp-up and L-mode operation into H-mode and high-performance flattop conditions. More importantly, because the transition and pedestal models are embedded in the same differentiable graph as the transport, heating, current, and geometry modules, TokaGrad can optimize not only the final H-mode state, but also the dynamical pathway required to access it.

\section{Interactive live simulation}\label{sec3}

A practical advantage of the JAX implementation is that the simulation can benefit from just-in-time (JIT) compilation and hardware acceleration on GPUs or TPUs. Once compiled, repeated evaluations of the transport step and the full scenario trajectory can be executed efficiently. In addition, because TokaGrad is designed to preserve differentiability, it naturally relies on accelerated differentiable components, including NN surrogates for transport, neoclassical, and pedestal closures. This combination of JIT-compiled solvers and fast surrogate closures enables faster-than-real-time simulation, making it suitable not only for offline scenario analysis but also for repeated full-scenario evaluations during optimization. Such speed is particularly important for gradient-based workflows, where many forward and backward evaluations are required to optimize actuator trajectories, scenario parameters, or reactor-design variables.

To make use of this capability, TokaGrad includes an interactive live-simulation user interface, which can be used for educational and practical purposes. The interface allows users to modify actuator and scenario parameters while observing the corresponding plasma response. The simulator can be run under fixed design and actuator conditions, but it also supports time-dependent scenario waveforms, including ramp-up, flattop, and ramp-down phases. Actuator quantities such as plasma current, toroidal magnetic field, auxiliary heating power, density target, and shape parameters can be controlled as time-dependent waveforms. This interactive capability is useful for exploring intuitive cause-and-effect relationships, testing actuator authority, and developing control strategies before applying more formal optimization. For example, changes in auxiliary heating can be observed through the subsequent temperature rise, alpha-heating feedback, L-H transition gate, pedestal formation, and fusion-power response.

\begin{figure}[t!]
\centering
\includegraphics[width=\linewidth]{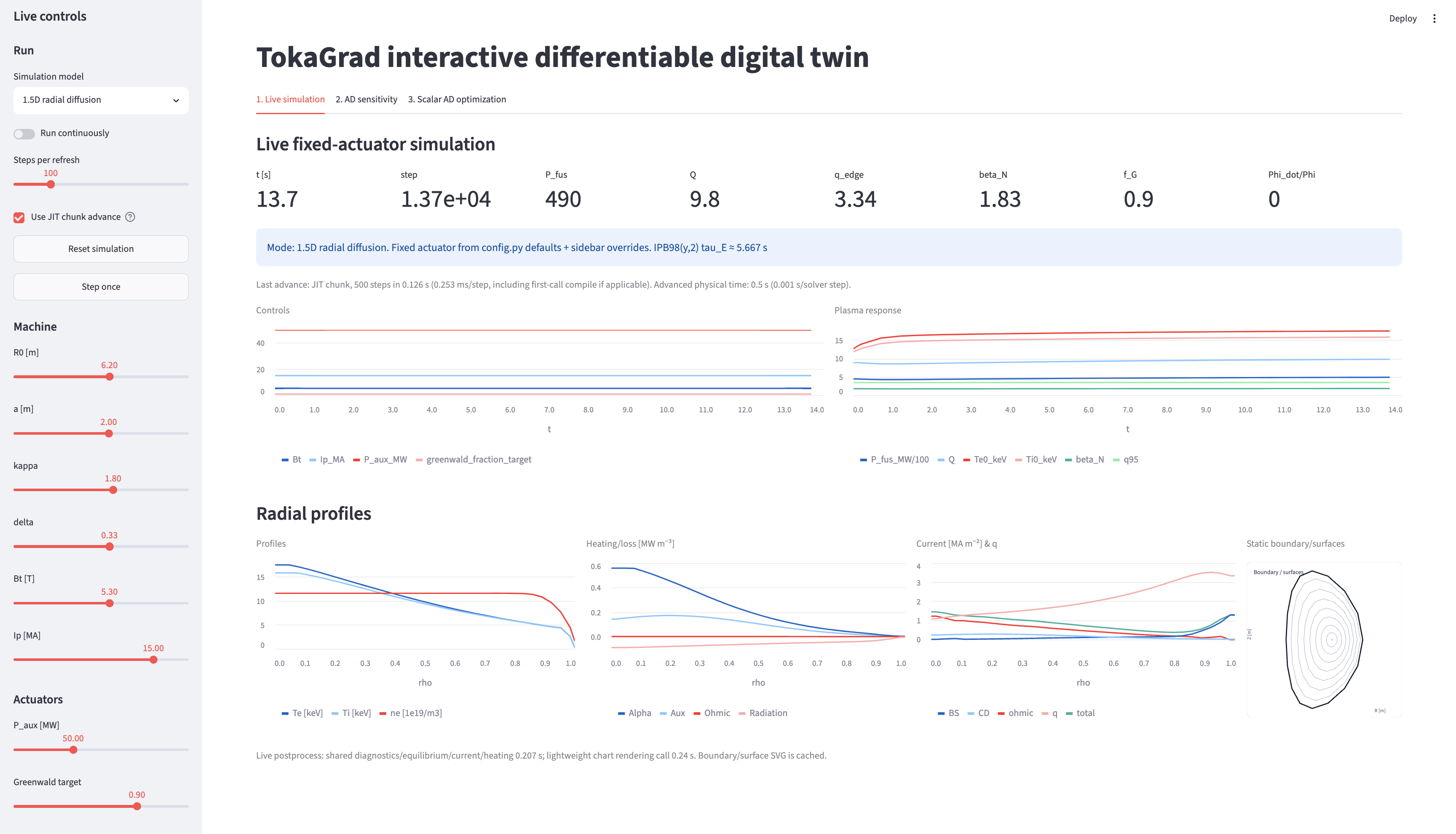}
\caption{A graphical user interface for interactive simulation with TokaGrad. The sidebar on the left includes actuator controls and scenario settings, and the right shows the corresponding plasma responses.}\label{fig2}
\end{figure}

Figure \ref{fig2} shows the graphical user interface developed for interactive TokaGrad simulations. The interface displays actuator controls and scenario settings together with time-evolving plasma profiles and scalar performance metrics. Typical outputs include electron and ion temperature, density, heating and radiation terms, current-density components, safety factor, fusion power, and fusion gain. By combining forward simulation, live actuator adjustment, and immediate visualization of the plasma response, the interface provides an educational and practical environment for scenario exploration and for diagnosing the behavior of the integrated differentiable model.

\subsection{Benchmarks}\label{sec3.1}

Because an integrated transport simulator couples many physics modules through their inputs and outputs, small inconsistencies in one module can propagate through the full simulation and lead to substantial differences in the final plasma state. For this reason, validating each module in isolation is not sufficient. A meaningful test requires comparison at the integrated-simulation level, where the feedback among transport, sources, current evolution, and geometry is active. To assess the consistency of TokaGrad with existing differentiable tokamak transport modeling, we perform a benchmark against TORAX \cite{citrin2024_torax}, a JAX-based differentiable transport simulator.

\begin{figure}[t!]
\centering
\includegraphics[width=\linewidth]{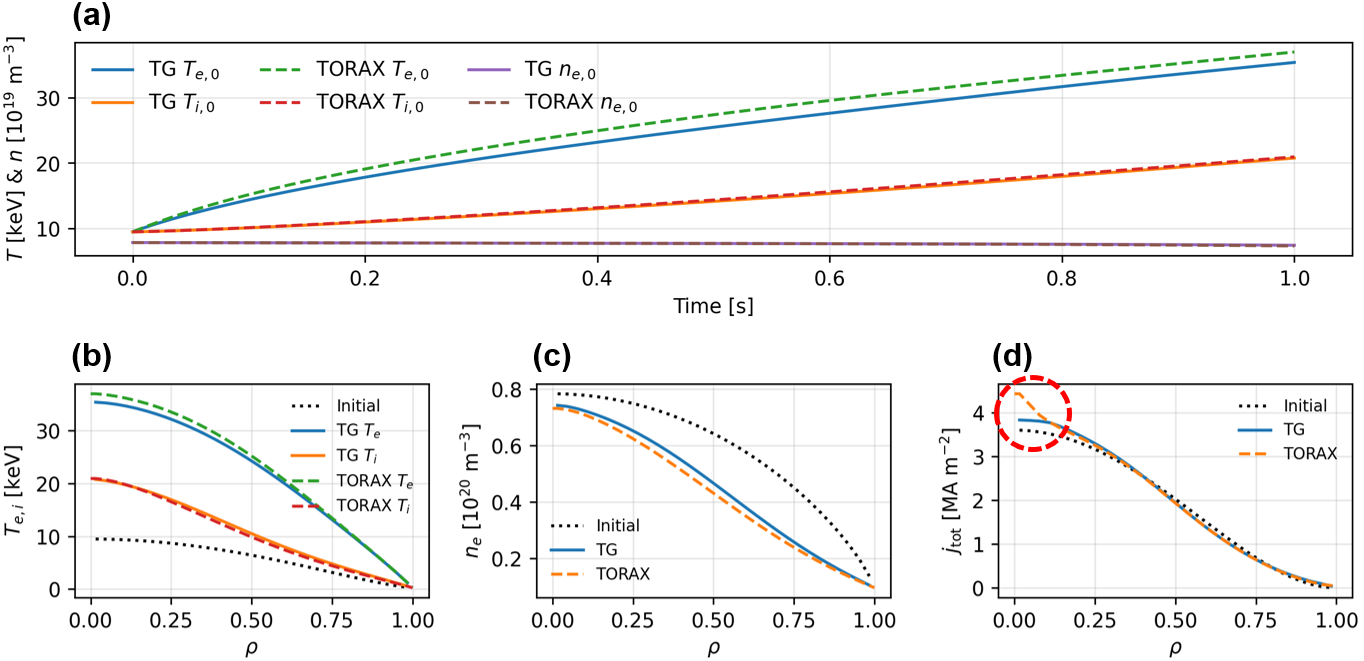}
\caption{Comparison of the simulation results from TokaGrad (TG) and TORAX. The TokaGrad results are shown in solid lines and the TORAX results are in dashed lines. \textbf{a}. Time evolution of the central values of the electron and ion temperatures ($T_{\rm e,i,0}$) and electron density ($n_{\rm e,0}$). \textbf{b}. Final electron ($T_{\rm e}$) and ion ($T_{\rm i}$) temperature profiles, where the excessively high $T_{\rm e}$ is due to artificially high ohmic heating caused by plasma resistivity increased 100-fold. \textbf{c}. Final electron density profiles. \textbf{d}. Final total current density profiles. The initial profiles for temperature, density, and current density are also shown in black dotted lines.}\label{fig3}
\end{figure}

Figure \ref{fig3} compares the simulation results obtained from TORAX (dashed) and TokaGrad (solid) under matched conditions. Both simulations use an ITER-like ($I_{\rm p}=15 \text{ MA}, B_{\rm T}=5.3 \text{ T}, R_0=6.2 \text{ m}, a=2.0 \text{ m}$) circular geometry and a Gaussian auxiliary-heating deposition profile with a 20 MW heating power. The intrinsic heating includes ohmic heating, alpha heating, electron-ion heat exchange, and radiation power loss. The transport closure is chosen to be the BgB model, which is common in both frameworks. The boundary conditions, initial profiles, and machine parameters are matched as closely as possible. The comparison is performed for an L-mode scenario because TORAX currently does not include the same differentiable L-H transition and dynamic pedestal treatment used in TokaGrad. The benchmark therefore tests whether TokaGrad reproduces the expected core-transport response before the additional H-mode and pedestal modules are activated. The number of radial grid is $N_\rho=32$ and the timestep is $\Delta t=0.001 \text{ s}$. Here, the plasma resistivity is intentionally multiplied by 100 to see meaningful poloidal flux diffusion over a short period.

The two simulations show similar time evolution and final profile structure for the main plasma quantities. In particular, the electron and ion temperature profiles, density profile, and current density profile remain within the same qualitative range within 5\% error when the same transport and source assumptions are used. The excessively high $T_{\rm e}$ profiles in the two simulations are due to artificially high ohmic heating caused by plasma resistivity increased 100-fold. Although an error of about 5\% occurs in the core $T_{\rm e}$, this is presumed to be due to the current density bump in the center shown in Figure \ref{fig3}d with a dashed circle, which is also excessively influenced by the 100-fold increase in resistivity. Since there is no bootstrap or central current drive and current diffusion in the core is low, this bump is considered a numerical artifact caused by the core boundary condition for $\psi$. The agreement in the two simulations indicates that the differentiable transport backbone of TokaGrad is consistent with an established JAX-based transport framework in a simplified L-mode setting.

We also compared the wall-clock time as a function of the simulated discharge pulse length. This comparison is important for ITER-scale simulations, where discharge durations of several hundred seconds may be required, and for differentiable optimization, where many repeated forward and backward evaluations are needed. To accelerate long-pulse and repeated simulations, TokaGrad uses JAX-based just-in-time (JIT) compilation, in which the numerical update graph is traced and compiled before execution. Once compiled, the same simulation kernel can be repeatedly evaluated with reduced Python overhead and optimized low-level operations.

\begin{figure}[t!]
\centering
\includegraphics[width=3in]{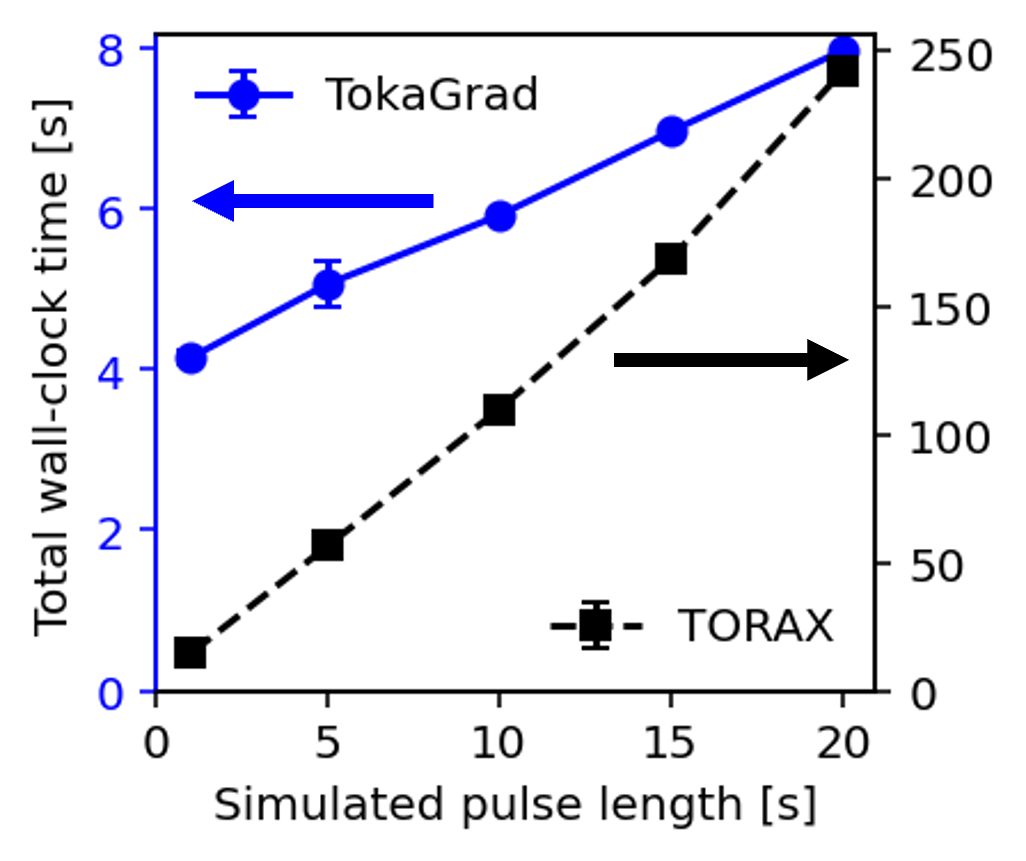}
\caption{Comparison of the wall-clock time of the TokaGrad and TORAX simulations. Under the same CPU environment, TokaGrad shows a more than one order of magnitude faster speed than TORAX. Here, the timestep in the simulations is fixed as $\Delta t = 0.001$ s.}\label{fig4}
\end{figure}

For a fair comparison, both simulations were performed with the semi-implicit numerical scheme, in a JAX-enabled setting on the same CPU environment using an Apple M1 Pro. Other physical and numerical settings are the same as the simulations in Figure \ref{fig3}. As shown in Figure \ref{fig4}, the wall-clock time of TORAX increases approximately in proportion to the simulated discharge duration, as is typical of conventional tokamak transport simulations. In contrast, TokaGrad exhibits a finite offset corresponding to the initial JIT compilation time, rather than a purely proportional scaling from the start. Once this compilation cost is excluded, however, the simulation advances substantially faster than TORAX, by more than one order of magnitude. For short-pulse simulations, the compilation overhead can therefore make TokaGrad less efficient. For long-pulse scenarios and repeated simulation tasks, however, this overhead is amortized, making TokaGrad particularly advantageous and well suited to differentiable optimization workflows. This allows a faster-than-real-time simulation for a pulse longer than $5\text{ s}$. We note that the absolute runtime of JAX-based simulations can differ significantly on GPU or TPU hardware.


In the benchmark against TORAX shown in Figures \ref{fig3} and \ref{fig4}, we used circular geometry, the empirical Bohm-gyroBohm transport model, and L-mode plasma conditions in order to match the model assumptions and simulation settings as closely as possible. However, as discussed in Section \ref{sec2}, TokaGrad also provides differentiable implementations of various equilibrium options, transport models, dynamic L-H transition, and pedestal physics. These capabilities enable more realistic prediction of H-mode plasmas.

\begin{figure}[t!]
\centering
\includegraphics[width=\linewidth]{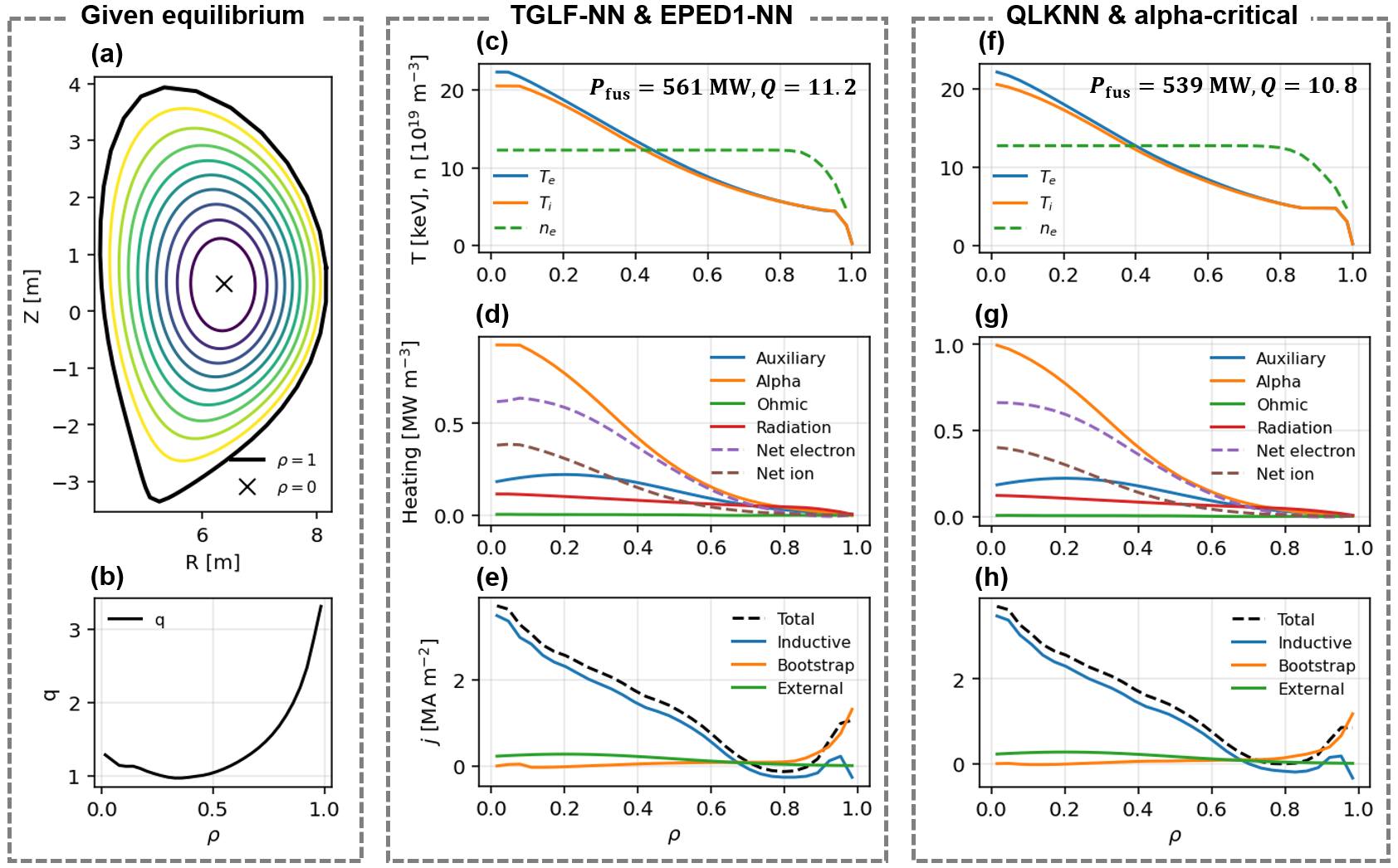}
\caption{Simulation results for H-mode plasmas under the ITER baseline scenario, using two different combinations of physics models in TokaGrad. \textbf{a}. The given EQDSK equilibrium. \textbf{b}. The $q$ profile from the equilibrium, while it can also be obtained from the current density profile by option. \textbf{c--e}. The plasma temperature, density, heating, and current density profiles simulated via TGLF-NN and EPED1-NN models. \textbf{f-h}. The profiles simulated via QLKNN and alpha-critical models.}\label{fig5}
\end{figure}

Figure \ref{fig5} shows simulation results for H-mode plasmas sustained by 50 MW of auxiliary heating under the ITER baseline scenario, using two different combinations of physics models. In the first case, shown in Figures \ref{fig5}c-e, turbulent transport and pedestal structure are modeled using TGLF-NN and EPED1-NN. In the second case, shown in Figures \ref{fig5}f-h, QLKNN and the alpha-critical pedestal model are used. In both cases, the equilibrium is fixed to the ITER-baseline EQDSK shown in Figure \ref{fig5}a, and the neoclassical model is given by NEO-NN. Particle transport (Equation \ref{eq_density_diffusion}), electron and ion heat transport (Equation \ref{eq_temperature_diffusion}), and poloidal flux diffusion (Equation \ref{eq_psi_diffusion}) were all solved.  Here, the $q$ profile (Figure \ref{fig5}b) is obtained from the given EQDSK, but it can be reconstructed from the current density as well by the user's option.

Both simulations show the plasma state after 10 s of relaxation under fixed actuator and machine conditions, while the electron density is feedback-controlled to match a prescribed Greenwald density fraction of 0.9. Although quantitative differences arise from the different model combinations, both cases produce well-known ITER-like temperature and density profiles, with a fusion power of approximately $\sim550$ MW and $Q \sim 11$. Even when high-fidelity surrogate models such as TGLF-NN, QLKNN, EPED1-NN, and NEO-NN are used, the wall-clock time for a 10 s simulation is approximately 7 s, demonstrating faster-than-real-time simulation capability.

\subsection{Full-discharge simulation of the ITER L-to-H scenario}\label{sec3.2}

In Figure \ref{fig5}, we simulated H-mode sustainment during the flattop phase of a 50 MW ITER baseline scenario. However, whether the plasma can reach this state starting from the initial ramp-up phase is a different question. In particular, this requires assessing whether the plasma can transition from L-mode to H-mode. In ITER, approximately 50-73 MW of auxiliary heating power is expected to be available, while the predicted L-H transition threshold is also in this range \cite{Martin_2008_threshold, Delabie_2026_threshold}. Therefore, to achieve an effective L-H transition, it is necessary not only to adjust the fixed actuator conditions during the flattop phase, but also to optimize the current and heating ramp-up waveforms so that positive feedback effects, such as alpha heating, can be utilized as much as possible.

\begin{figure}[t!]
\centering
\includegraphics[width=\linewidth]{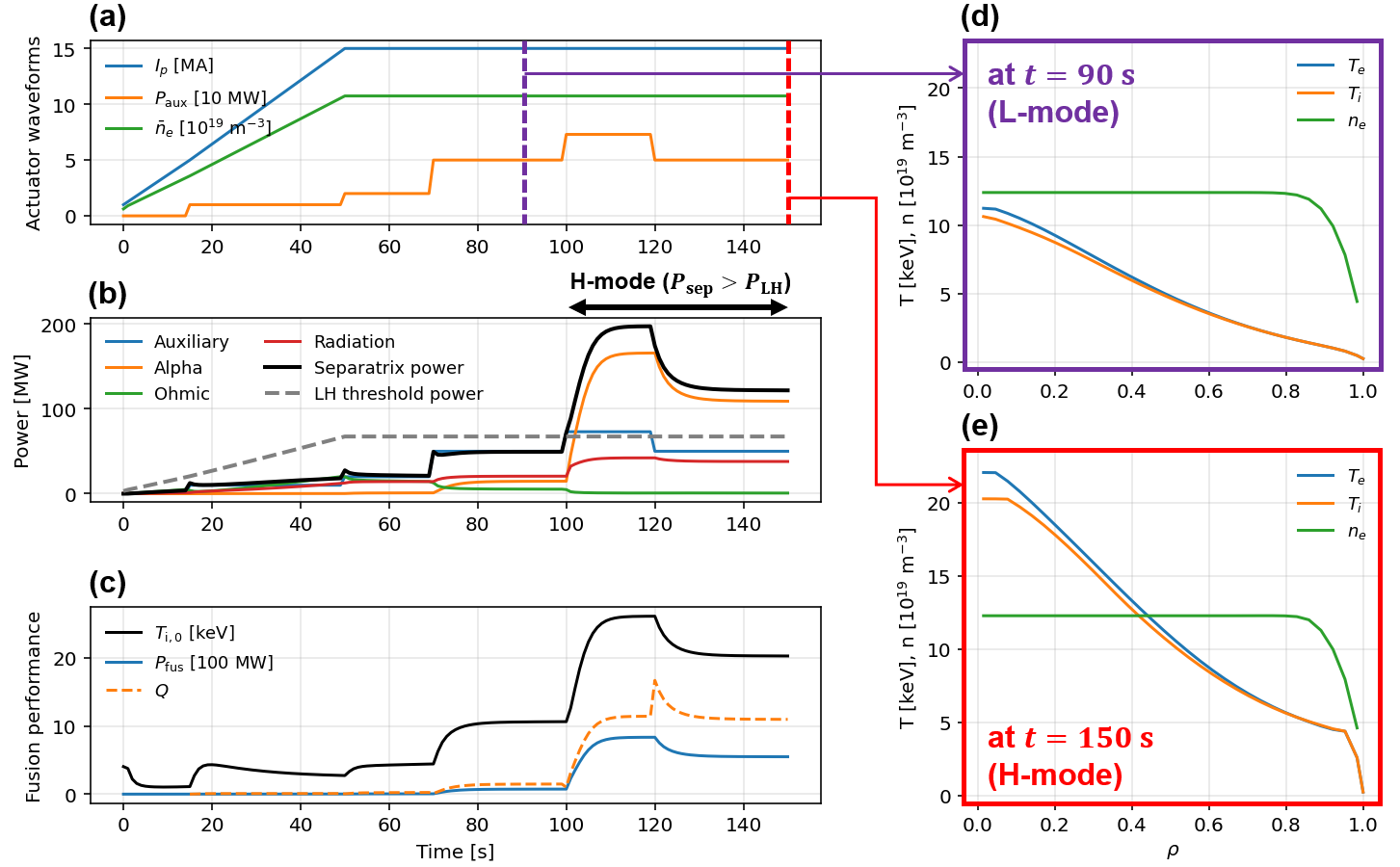}
\caption{TokaGrad simulation of the L- to H-mode full ITER baseline scenario. \textbf{a}. Waveforms of the plasma current, auxiliary heating, and the averaged density. \textbf{b}. Time evolution of heating, power loss, and the L-H threshold power. \textbf{c}. Time evolution of the central ion temperature, fusion power, and fusion gain. \textbf{d, e}. Temperature and density profiles at $t=90\text{ s}$ and $t=150\text{ s}$, which are under the same actuator condition.}\label{fig6}
\end{figure}

In this study, we performed a TokaGrad simulation of the ITER baseline scenario using the actuator waveforms shown in Figure \ref{fig6}a, in which the heating power is transiently overshot. Starting from the ramp-up phase, we computed the transition and subsequent saturation process. The final actuator conditions, including heating power and plasma current, are identical to those used in Figure \ref{fig5}. The transport models are TGLF-NN and NEO-NN, the pedestal model is EPED1-NN, and the equilibrium is prescribed using the EQDSK corresponding to the ITER baseline scenario. The occurrence of the L-H transition is determined based on the Delabie threshold scaling \cite{Delabie_2026_threshold}. Except for the time-dependent actuator waveforms, the simulation settings are the same as those used in Figure \ref{fig5}. The simulation of this 150 s discharge took a wall-clock time of about 17 s.

In Figure \ref{fig6}a, as the plasma current ramps up, the density increases to maintain the prescribed Greenwald fraction. When auxiliary heating is turned on, the temperature also rises, which in turn enhances the fusion reaction rate and alpha heating. However, at $t=90$ s, when the auxiliary heating has reached 50 MW (same as Figure \ref{fig5}), the separatrix power, the sum of heating and loss, remains below the L-H transition threshold, and the plasma retains an L-mode profile, as shown in Figure \ref{fig6}d. Because of the relatively low temperature, the fusion power and $Q$ are also much lower than those in Figure \ref{fig5}.

After the auxiliary heating is transiently increased to 73 MW at $t=100$ s, the total heating power exceeds the threshold and the plasma transitions to H-mode. Once the transition occurs, the increase in temperature enhances the fusion reaction rate, and the resulting alpha heating remains large enough to keep the total heating above the threshold even after the auxiliary heating is reduced back to 50 MW. This behavior demonstrates hysteresis in the L-H transition and H-mode sustainment. As a result, after sufficient relaxation time, the plasma achieves a fusion power above 500 MW and $Q>10$. This result is consistent with previous integrated-simulation studies of the ITER baseline scenario \cite{SHKim_2018_heating_overshoot_transition}.

\section{Optimization with automatic differentiation}\label{sec4}

The greatest advantage of differentiable simulation is that gradients can propagate among the parameters within the simulation. These parameters include design variables, time-dependent actuator controls, and plasma responses. This makes it possible to compute Jacobians that describe how plasma performance at a later time is determined by design choices or control inputs. Such gradients can also be used for gradient-based optimization, which is widely employed in deep learning. Through this approach, one can search for optimal design or control conditions that achieve a desired plasma state. In this section, we present examples in which TokaGrad combines differentiable simulation with gradient-based optimization and applies it to actuator control, scenario optimization, and reactor design. Although the resulting optimized solution may depend on the selected physics models, this framework provides a promising new methodology for avoiding conventional trial-and-error-based scenario optimization workflows.

\subsection{Actuator control}\label{sec4.1}

\begin{figure}[t!]
\centering
\includegraphics[width=\linewidth]{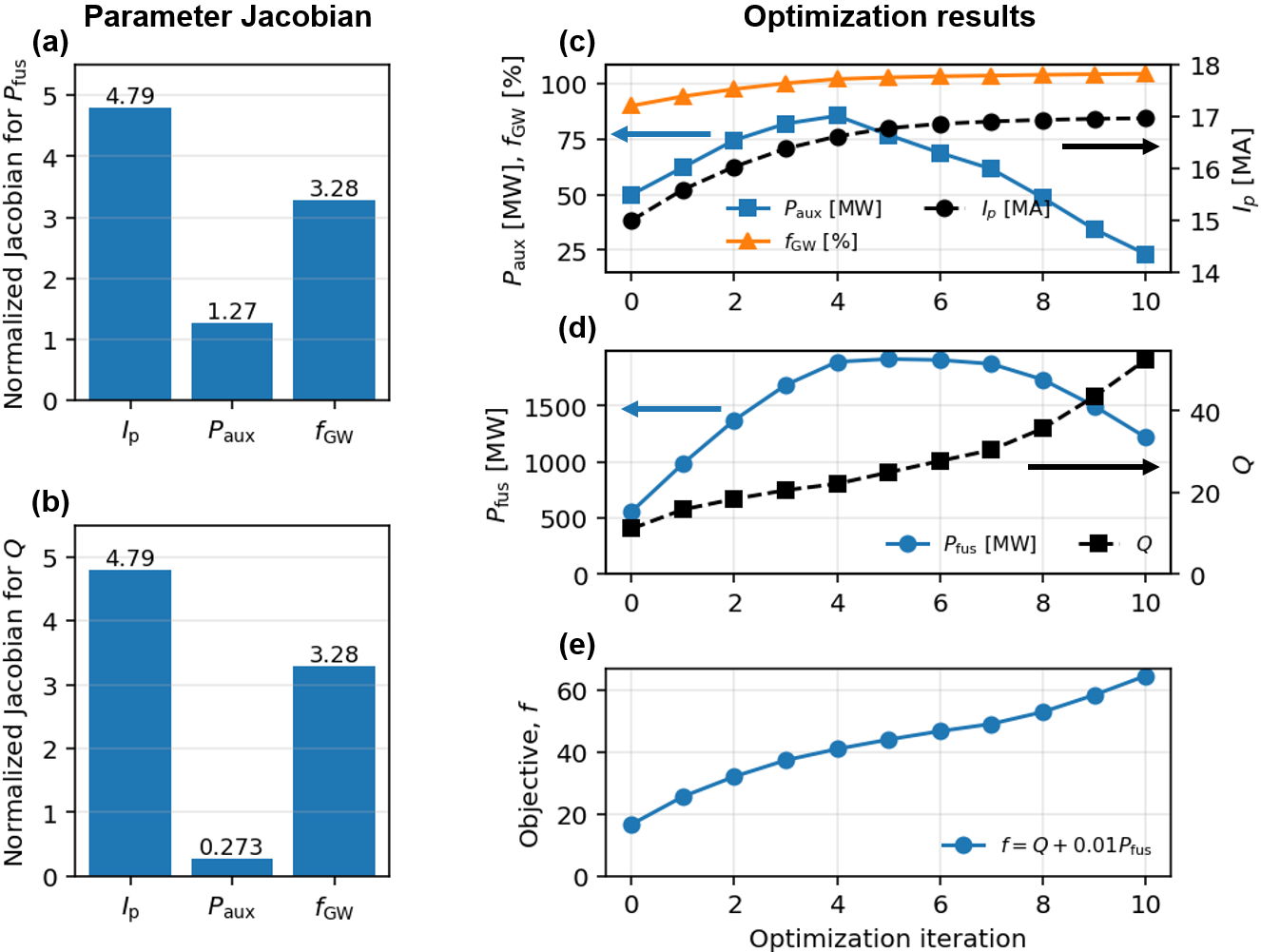}
\caption{Actuator dependency on fusion metrics and gradient-based actuator control. \textbf{a, b}. Normalized Jacobian ($\partial \ln{y}/\partial \ln{x}$) for fusion power and gain with respect to the actuators. \textbf{c}. Gradient-based control of the actuators to maximize the fusion metrics. \textbf{d}. Resulting fusion metrics according to the controlled actuators. \textbf{e}. Resulting objective function consisting of the weighted sum of fusion power and gain.}\label{fig7}
\end{figure}

A JAX-based framework enables the computation of Jacobians among simulation parameters through automatic differentiation. In a differentiable simulation, the computational graph is connected from the initial condition all the way to the final plasma profile. Therefore, it can provide not only the dependency after a single time step, but also the dependency of the nonlinearly evolved plasma response. Figures \ref{fig7}a and \ref{fig7}b show the normalized Jacobians of $P_{\rm fus}$ and $Q$ after 10 s of relaxation with respect to actuator parameters such as $P_{\rm aux}$, $I_{\rm p}$, and $f_{\rm GW}$. The simulation setting is the same as in Figure \ref{fig5}, which uses the TGLF-NN and EPED1-NN models. Here, the normalized Jacobian is defined as $\partial \ln{y}/\partial \ln{x}$, where $y \in P_{\rm fus}(t=10\text{ s}), Q(t=10\text{ s})$ and $x \in P_{\rm aux}, I_{\rm p}, f_{\rm GW}$. For example, they provide the sensitivity of fusion power to changes in $P_{\rm aux}$. The trend seen in Figures \ref{fig7}a and \ref{fig7}b, namely that increasing $P_{\rm aux}$ meaningfully increases $P_{\rm fus}$ while producing only a small change in $Q$, is also consistent with the well-known physics of the ITER baseline scenario.

Knowing the gradients among these parameters also means that, conversely, the actuators can be controlled so as to steer the plasma toward a desired state. In this study, the objective function is defined as $f=(Q + 0.01 P_{\rm fus})$ at $t=10\text{ s}$, and the Adam optimizer (learning rate 1.0) \cite{kingma2017_adam} implemented in the Optax package \cite{deepmind2020_jax_optax} is used. The allowable ranges of the control parameters are set to $[2\text{ MA},17\text{ MA}]$, $[0\text{ MW},120\text{ MW}]$, and $[0.5,1.05]$ for $I_p$, $P_{\rm aux}$, and $f_{\rm GW}$, respectively. In addition, in order to avoid plasma instabilities that may arise during actuator control, the constraints $q_{95} > 3$ and $\beta_N < 3$ are imposed. These constraints are implemented as penalty terms added to the objective function.

Figures \ref{fig7}c and \ref{fig7}d show the results of controlling the actuators using a gradient-ascent algorithm based on this gradient information so as to increase $P_{\rm fus}$ and $Q$. Here, the first iteration took 70 seconds due to JIT precompilation, and the remaining iterations each took about 23 seconds, on an Apple M1 Pro CPU. During the early optimization iterations, the actuators increase gradually along the positive gradient direction shown in Figures \ref{fig7}a and \ref{fig7}b, and accordingly both $P_{\rm fus}$ and $Q$ increase. However, because the dependency among parameters is non-monotonic, one can see that $P_{\rm aux}$ is first increased and then reduced again in order to maximize the objective, $Q+0.01 P_{\rm fus}$. This can be interpreted as follows: as the iterations proceed, $I_p$ and $f_{\rm GW}$ increase, causing the density to rise by about a factor of 1.4, which in turn leads to nearly a twofold improvement in fusion power ($\propto n_{\rm e}^2$). In this regime, alpha heating becomes dominant, so increasing $P_{\rm aux}$ no longer provides much benefit. Instead, reducing $P_{\rm aux}$ becomes a strategy for improving $Q$. Although the fusion power slightly decreases as $P_{\rm aux}$ is lowered, the summed objective in Figure \ref{fig7}e continues to increase due to strong $Q$ dependency. Whether this solution is practically feasible is another problem, but the results show that an actuator control solution based on AD can be obtained in a way that increases the objective in a given simulation. Of course, this result is obtained under the given physics models and initial conditions, and the optimal control outcome can vary depending on the models, constraints employed, and the objective functions.

\subsection{Scenario optimization}\label{sec4.2}

\begin{figure}[t!]
\centering
\includegraphics[width=\linewidth]{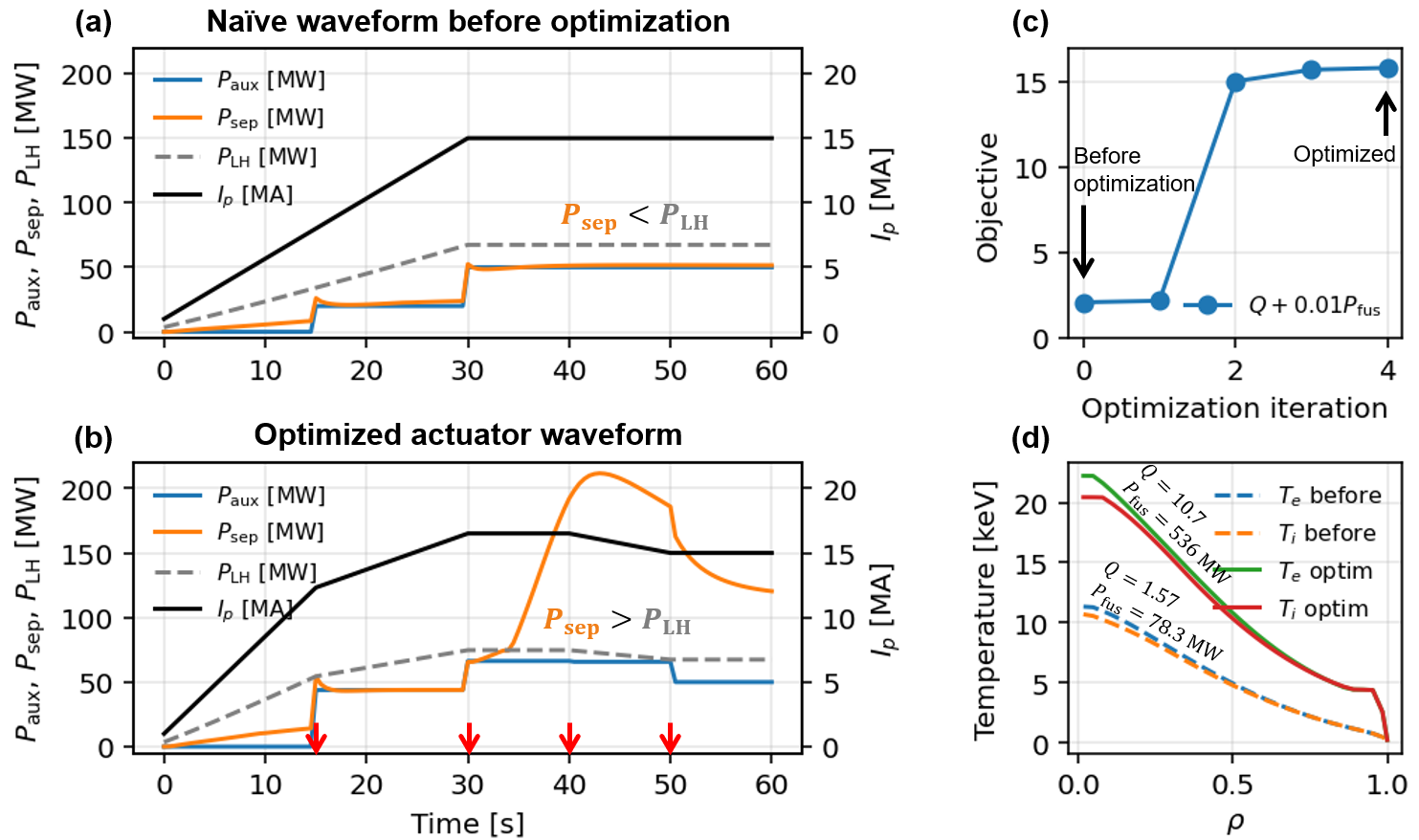}
\caption{Operation scenario waveform optimization for effective L-H transition. \textbf{a}. A naive plasma current and heating ramp-up scenario, which results in a low fusion performance with an L-mode plasma. \textbf{b}. Final optimized ramp-up scenario, which results in a high fusion performance with an H-mode plasma. \textbf{c}. Evolution of the objective function with respect to optimization iterations. \textbf{d}. Comparison of the temperature profiles between before and after optimization.}\label{fig8}
\end{figure}

As shown in Figures \ref{fig6}d and \ref{fig6}e, even under the same actuator conditions in the ITER baseline scenario, whether the plasma can transition to H-mode depends on the scenario waveform from ramp-up to flattop. This is because the available auxiliary heating power in ITER is below, or comparable to, the L-H transition threshold. Indeed, in a naive ramp-up scenario shown in Figure \ref{fig8}a, the power crossing the plasma separatrix, $P_{\rm sep}$, does not exceed the L-H threshold, resulting in L-mode operation and low fusion performance, as indicated by the dashed line in Figure \ref{fig8}d.

However, the L-H transition exhibits hysteresis due to the nonlinearity of turbulent transport and feedback from alpha heating. Therefore, with an appropriately designed ramp-up scenario for the plasma current and auxiliary heating, it is possible to reach H-mode even under the same final actuator conditions. In this section, we start from a naive scenario waveform in Figure \ref{fig8}a and optimize the waveform to increase the fusion power and fusion gain, testing whether the optimization can recover a heating-overshoot solution similar to that shown in Figure \ref{fig6}. For the optimization process, unlike the single waveform simulation in Figure \ref{fig6}, multiple iterations including forward and backward evaluations of simulations with the same settings are required. Therefore, the ramp-up and flat-top timelines were reduced for proof-of-concept, and the timestep $\Delta t$ was increased to 0.01 s. Unlike the fixed-actuator optimization in Section \ref{sec4.1}, this corresponds to optimizing the actuator values at multiple time points. In this case, we already know that transient power overshooting is required to access H-mode in the ITER baseline scenario. Nevertheless, demonstrating that the optimizer can recover this solution provides a proof of concept for future applications to advanced scenario development aimed at achieving higher fusion performance.

In the TokaGrad optimization, the initial conditions, $I_p=1\text{ MA}$ and $P_{\rm aux}=0\text{ MW}$, and the final conditions, $I_p=15\text{ MA}$ and $P_{\rm aux}=50\text{ MW}$, are fixed to match the ITER baseline scenario. The waveform between these initial and final conditions is then parameterized by four intermediate time points, and gradient-based optimization is performed by controlling the actuator values at those points. These four selected time points are indicated by the vertical red arrows in Figure \ref{fig8}b. To simplify the problem, the Greenwald fraction is fixed at 0.9, so that the density varies proportionally with the plasma current. To avoid unrealistic waveforms, the controllable ranges of $I_p$ and $P_{\rm aux}$ are set to $[5\text{ MA},17\text{ MA}]$ and $[5\text{ MW},73\text{ MW}]$, respectively. As in Section \ref{sec4.1}, the stability constraints $q_{95}>3$ and $\beta_N<3$ are also imposed.

Figure \ref{fig8}c shows the evolution of the objective function, $Q+0.01P_{\rm fus}$, over the optimization iterations. The jump that occurs around iteration 2 can be interpreted as a bifurcation associated with H-mode access enabled by waveform optimization. The final optimized waveform is shown in Figure \ref{fig8}b. As expected, the optimizer raises $P_{\rm aux}$ at an intermediate time point to enter H-mode, and subsequently lowers it. By transiently increasing $P_{\rm aux}$, the condition $P_{\rm sep}>P_{\rm LH}$ is satisfied at around $t=35$ s, triggering the L-H transition. After the transition, alpha-heating feedback allows $P_{\rm sep}>P_{\rm LH}$ to be maintained even when $P_{\rm aux}$ is reduced again. As a result, the plasma enters and sustains H-mode, as shown by the solid line in Figure \ref{fig8}d, and achieves the high fusion performance targeted in the ITER baseline scenario. Here, the first iteration took 67 seconds due to JIT precompilation, and the remaining iterations each took about 13 seconds. Although the present result is demonstrated for a known H-mode-access solution in the ITER baseline scenario as a proof of concept for scenario optimization, the same approach can be applied to advanced scenario development under different machine conditions and objectives, including hybrid and steady-state scenarios.

\subsection{Reactor design}\label{sec4.3}

The optimization studies in the previous sections correspond to controlling actuators, such as heating power and plasma current, under fixed device-design conditions. However, differentiable simulation and optimization can also be applied to reactor design. In reactor design, it is important not only to maximize the objective function, but also to properly account for engineering and physics constraints. Here, we perform an optimization experiment in which both actuator and design parameters are varied in order to achieve a high objective value under prescribed constraints.

Starting from the ITER baseline design, we optimize the design parameters $R_0$, $a$, $B_t$, $I_p$, and $P_{\rm aux}$ to maximize the objective function $Q+0.01P_{\rm fus}$. The plasma-stability constraints $q_{95}>3$ and $\beta_N<3$ are also included in the objective as penalty terms. Unlike the simulations and optimizations in the previous sections, this case does not use a fixed EQDSK equilibrium. Instead, a Miller/moment-based fixed-boundary equilibrium solver is used, because the shape parameters vary at each optimization iteration. The physics models are the same as before, based on TGLF-NN and EPED1-NN.

\begin{figure}[t!]
\centering
\includegraphics[width=4.5in]{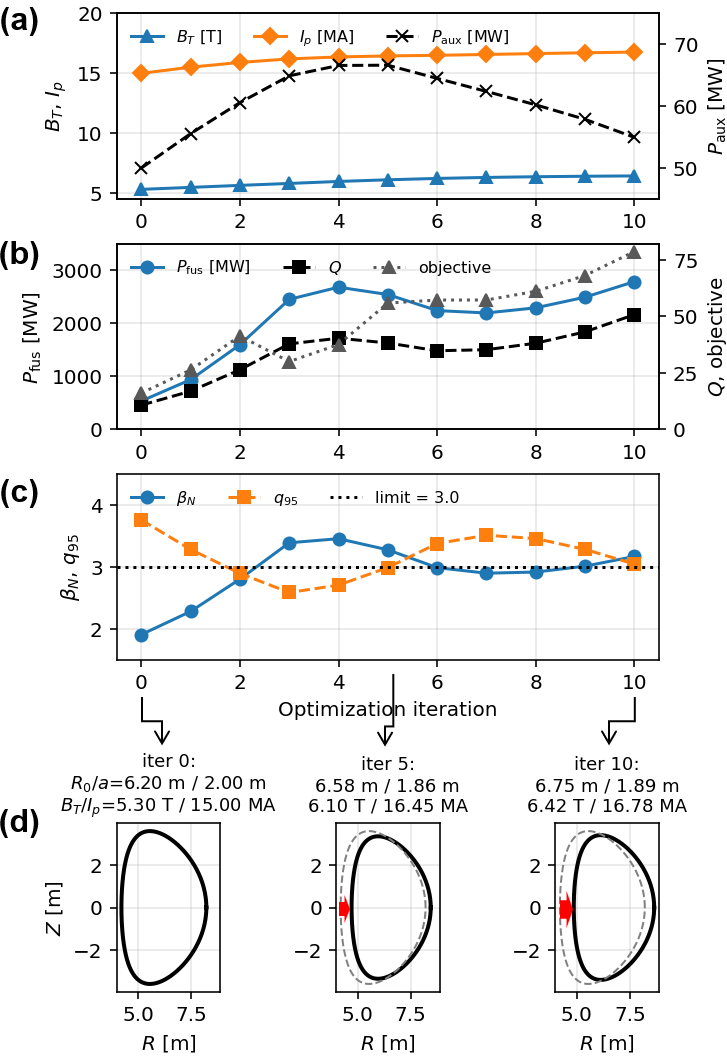}
\caption{Reactor design optimization with TokaGrad. \textbf{a}. Change of the design variables with respect to the optimization iteration. \textbf{b}. Change of the fusion performance. \textbf{c}. Change of the stability-related parameters. \textbf{d}. Change of the plasma shape parameters.}\label{fig9}
\end{figure}

Figure \ref{fig9} shows the evolution of the design parameters and plasma performance over the optimization iterations. As shown in Figure \ref{fig9}a and \ref{fig9}d, the optimizer increases the major radius, plasma current, and magnetic field strength, while slightly reducing the minor radius toward a higher-aspect-ratio configuration. Empirically, these changes are indeed favorable for confinement \cite{Cordey_2005_itpa04_scaling}, and accordingly both fusion power and fusion gain increase, as shown in Figure \ref{fig9}b. The auxiliary heating power initially increases slightly, but once the plasma enters a regime where alpha heating becomes dominant, it tends to decrease again, which is the same as the results in Figure \ref{fig7}.

However, increasing the plasma current reduces $q_{95}$, and as the plasma performance improves, $\beta_N$ can exceed 3, potentially making the plasma unstable. In particular, at iteration 3, both $\beta_N$ and $q_{95}$ excessively violate their limits, as shown in Figure \ref{fig9}c, and the resulting penalty term lowers the objective function value shown in Figure \ref{fig9}b. After this point, the design parameters are adjusted so as to improve fusion performance while keeping the $\beta_N$ and $q_{95}$ constraints close to their marginal limits.

Figure \ref{fig9}d shows the difference in plasma shape between the initial ITER baseline configuration and the final optimized design. Consistent with well-known expectations, increasing the major radius is beneficial for fusion performance. However, it should be noted that this optimization does not include engineering-stability or cost constraints associated with increasing the device size or magnetic-field strength. In the future, incorporating such additional constraints is expected to enable higher-fidelity reactor-design optimization. Nevertheless, these results demonstrate that differentiable simulation combined with gradient-based optimization can obtain meaningful reactor-design improvements with far fewer iterations than conventional random-search-based system-code approaches. Here, the first iteration took 152 seconds due to JIT compilation, and the remaining iterations each took about 14 seconds, on an Apple M1 Pro CPU.

\section{Conclusion}\label{sec5}

In this work, we developed ``TokaGrad'', an end-to-end differentiable tokamak transport simulation framework that covers the full plasma scenario from ramp-up to flattop, from core to edge, and across the L-to-H transition. To our knowledge, this is the first simulation and optimization framework that differentiably models dynamically evolving tokamak scenarios including equilibrium evolution, pedestal formation/sustainment, and abrupt confinement-regime transitions such as the L-to-H transition. As a result, gradients can propagate from the final plasma performance back to machine-design parameters, actuator controls, and initial plasma conditions through the entire simulation trajectory.

TokaGrad combines differentiable physics modules with JAX-based just-in-time compilation, enabling efficient forward and backward evaluations. Even when high-fidelity surrogate models such as TGLF-NN, EPED1-NN, and NEO-NN are employed, TokaGrad achieves faster-than-real-time simulation in the examples considered here. This computational speed, together with automatic differentiation, makes it possible to perform multiple iterations of forward and backward simulation even on laptop-scale hardware. We demonstrated this capability through fixed-actuator control, scenario-waveform optimization, and reactor-design optimization. A particularly important application is ITER baseline scenario optimization. In ITER, the available auxiliary heating power is expected to be marginal for L-H access, making the transition sensitive to the ramp-up waveform and to alpha-heating feedback. Using the fully differentiable L-to-H modeling capability of TokaGrad, we showed that gradient-based waveform optimization can recover a heating-overshoot strategy that enables effective H-mode access and sustainment under the same final actuator conditions. This demonstrates that differentiable scenario optimization can go beyond static flattop analysis and directly address path-dependent plasma evolution.

Although the demonstrations in this study focused on relatively simple optimization problems under ITER-baseline-like conditions, the framework is general. The same approach can be extended to gradient-based plasma control, advanced scenario development, and next-generation reactor optimization. In particular, future work can incorporate more complete engineering constraints, uncertainty quantification, multi-objective optimization, and experimental validation against time-dependent tokamak discharges. By enabling gradients to connect reactor design, actuator trajectories, and fusion performance control, TokaGrad provides a new route toward automated and physics-informed optimization of burning-plasma scenarios.

\backmatter

\section*{Acknowledgements}
This work was supported by a National Research Foundation of Korea (NRF) grant funded by the Korea government (MSIT) (Grant No. RS-2024-00346024).

\section*{Author contributions}
J.S. contributed to the design of this study, code development, numerical experiments, and writing the manuscript.

\section*{Data availability}
The codes that generate data in this paper will be available at the GitHub repository, https://github.com/jaem-seo/tokagrad.

\section*{Declarations}
The authors declare that they have no known competing financial interests or personal relationships that could have appeared to influence the work reported in this paper.

\begin{appendices}

\section{JAX conversion of surrogate models}\label{secA1}

Many high-fidelity surrogate models for tokamak physics, including transport, pedestal, and heating, have originally been trained and distributed in TensorFlow \cite{Abadi2026_tensorflow} or PyTorch \cite{NEURIPS2019_pytorch} formats. To integrate these models into TokaGrad while preserving end-to-end differentiability, we convert the trained neural-network surrogates into JAX-compatible functions. In practice, the trained weights and normalization constants are first extracted from the original model checkpoints. The corresponding network architecture is then reimplemented using JAX operations, with all affine layers, activation functions, residual connections, and input-output transformations reproduced explicitly. The extracted parameters are mapped layer by layer into JAX arrays and used as static model parameters inside pure functional inference routines. This procedure avoids calling TensorFlow or PyTorch inside the simulation loop, allowing the surrogate models to be traced by JAX, compiled with just-in-time compilation, vectorized over radial grids, and differentiated through automatic differentiation. As a result, neural-network closures such as TGLF-NN, EPED1-NN, QLKNN, and NEO-NN can be embedded directly into the differentiable computational graph of TokaGrad, enabling gradients to propagate through both the physics solver and the surrogate-model predictions. In the future, surrogate models for NBI and RF heating will also be integrated, enabling more flexible heating control and optimization.

\end{appendices}

\bibliography{sn-bibliography}

@article{KWON2020_kdemo,
title = {Recent progress in the design of the K-DEMO divertor},
journal = {Fusion Engineering and Design},
volume = {159},
pages = {111770},
year = {2020},
issn = {0920-3796},
doi = {https://doi.org/10.1016/j.fusengdes.2020.111770},
url = {https://www.sciencedirect.com/science/article/pii/S0920379620303185},
author = {Sungjin Kwon and Kihak Im and Suk-Ho Hong and Hyungho Lee and Thomas D. Rognlien and William Meyer and Keeman Kim},
}

@article{Rodriguez-Fernandez_2022_sparc,
doi = {10.1088/1741-4326/ac1654},
url = {https://doi.org/10.1088/1741-4326/ac1654},
year = {2022},
month = {mar},
publisher = {IOP Publishing},
volume = {62},
number = {4},
pages = {042003},
author = {Rodriguez-Fernandez, P. and Creely, A.J. and Greenwald, M.J. and Brunner, D. and Ballinger, S.B. and Chrobak, C.P. and Garnier, D.T. and Granetz, R. and Hartwig, Z.S. and Howard, N.T. and Hughes, J.W. and Irby, J.H. and Izzo, V.A. and Kuang, A.Q. and Lin, Y. and Marmar, E.S. and Mumgaard, R.T. and Rea, C. and Reinke, M.L. and Riccardo, V. and Rice, J.E. and Scott, S.D. and Sorbom, B.N. and Stillerman, J.A. and Sweeney, R. and Tinguely, R.A. and Whyte, D.G. and Wright, J.C. and Yuryev, D.V.},
title = {Overview of the SPARC physics basis towards the exploration of burning-plasma regimes in high-field, compact tokamaks},
journal = {Nuclear Fusion},
}

@article{McNamara_2024_st40,
doi = {10.1088/1741-4326/ad6ba7},
url = {https://doi.org/10.1088/1741-4326/ad6ba7},
year = {2024},
month = {aug},
publisher = {IOP Publishing},
volume = {64},
number = {11},
pages = {112020},
author = {McNamara, S.A.M. and Alieva, A. and Anastopoulos Tzanis, M.S. and Asunta, O. and Bland, J. and Bohlin, H. and Buxton, P.F. and Colgan, C. and Dnestrovskii, A. and du Toit, E. and Fontana, M. and Gemmell, M. and Gryaznevich, M.P. and Hakosalo, J. and Hardman, M.R. and Harryman, D. and Hoffman, D. and Iliasova, M. and Janhunen, S. and Janky, F. and Lister, J.B. and Lowe, H.F. and Maartensson, E. and Marsden, C. and Medvedev, S.Y. and Mirfayzi, S.R. and Moscheni, M. and Naylor, G. and Nemytov, V. and Njau, J. and O’Gorman, T. and Osin, D. and Pyragius, T. and Rengle, A. and Romanelli, M. and Romero, C. and Sertoli, M. and Shevchenko, V. and Sinha, J. and Sladkomedova, A. and Sridhar, S. and Stirling, J. and Takase, Y. and Thomas, P.R. and Varje, J. and Vekshina, E. and Vincent, B. and Willett, H.V. and Wood, J. and Wooldridge, E. and Zakhar, D. and Zhang, X. and Battaglia, D. and Bertelli, N. and Bonofiglo, P.J. and Delgado-Aparicio, L.F. and Duarte, V.N. and Gorelenkov, N.N. and de Haas, M. and Kaye, S.M. and Maingi, R. and Mueller, D. and Ono, M. and Podesta, M. and Ren, Y. and Trieu, S. and Delabie, E. and Gray, T.K. and Lomanowski, B. and Unterberg, E.A. and Marchuk, O. and the ST40 Team},
title = {Overview of recent results from the ST40 compact high-field spherical tokamak},
journal = {Nuclear Fusion},
}

@Article{Han2022_fire,
author={Han, H.
and Park, S. J.
and Sung, C.
and Kang, J.
and Lee, Y. H.
and Chung, J.
and Hahm, T. S.
and Kim, B.
and Park, J.-K.
and Bak, J. G.
and Cha, M. S.
and Choi, G. J.
and Choi, M. J.
and Gwak, J.
and Hahn, S. H.
and Jang, J.
and Lee, K. C.
and Kim, J. H.
and Kim, S. K.
and Kim, W. C.
and Ko, J.
and Ko, W. H.
and Lee, C. Y.
and Lee, J. H.
and Lee, J. K.
and Lee, J. P.
and Lee, K. D.
and Park, Y. S.
and Seo, J.
and Yang, S. M.
and Yoon, S. W.
and Na, Y.-S.},
title={A sustained high-temperature fusion plasma regime facilitated by fast ions},
journal={Nature},
year={2022},
month={Sep},
day={01},
volume={609},
number={7926},
pages={269-275},
issn={1476-4687},
doi={10.1038/s41586-022-05008-1},
url={https://doi.org/10.1038/s41586-022-05008-1}
}

@article{Na_2026_fire,
doi = {10.1088/1741-4326/ae332f},
url = {https://doi.org/10.1088/1741-4326/ae332f},
year = {2026},
month = {jan},
publisher = {IOP Publishing},
volume = {66},
number = {2},
pages = {026049},
author = {Na, Yong-Su and Park, S.J. and Han, H. and Lee, Janghyeok and Heo, C. and Hong, S.C. and Sung, C. and Kim, D. and Kang, J. and Lee, Y.H. and Chung, J. and Hahm, T.S. and Kim, B. and Bak, J.G. and Budny, R. and Cha, M.S. and Choi, G.J. and Choi, M.J. and Gwak, J. and Hahn, S.H. and Jang, J. and Lee, K.C. and Kim, J.H. and Kim, S.K. and Kim, W.C. and Ko, J. and Ko, W.H. and Lee, C.Y. and Lee, J.H. and Lee, J.H. and Lee, J.K. and Lee, J.P. and Lee, K.D. and Park, J.-K. and Park, J.M. and Park, Y.S. and Seo, J. and Yang, S.M. and Yoon, S.W. and KSTAR Team},
title = {On FIRE mode in KSTAR},
journal = {Nuclear Fusion},
}

@article{Snyder_2019_superH,
doi = {10.1088/1741-4326/ab235b},
url = {https://doi.org/10.1088/1741-4326/ab235b},
year = {2019},
month = {jun},
publisher = {IOP Publishing},
volume = {59},
number = {8},
pages = {086017},
author = {Snyder, P.B. and Hughes, J.W. and Osborne, T.H. and Paz-Soldan, C. and Solomon, W.M. and Knolker, M. and Eldon, D. and Evans, T. and Golfinopoulos, T. and Grierson, B.A. and Groebner, R.J. and Hubbard, A.E. and Kolemen, E. and LaBombard, B. and Laggner, F.M. and Meneghini, O. and Mordijck, S. and Petrie, T. and Scott, S. and Wang, H.Q. and Wilson, H.R. and Zhu, Y.B.},
title = {High fusion performance in Super H-mode experiments on Alcator C-Mod and DIII-D},
journal = {Nuclear Fusion},
}

@article{Tardini_2026_astra8,
doi = {10.1088/1361-6587/ae7640},
url = {https://doi.org/10.1088/1361-6587/ae7640},
year = {2026},
month = {jun},
publisher = {IOP Publishing},
volume = {68},
number = {6},
pages = {065024},
author = {Tardini, G and Fable, E and Angioni, C and Bergmann, M and Fajardo, D and Luda, T},
title = {ASTRA-8: a modern framework for transport analysis and modelling in fusion devices},
journal = {Plasma Physics and Controlled Fusion},
}

@article{PANKIN2025_transp,
title = {TRANSP integrated modeling code for interpretive and predictive analysis of tokamak plasmas},
journal = {Computer Physics Communications},
volume = {312},
pages = {109611},
year = {2025},
issn = {0010-4655},
doi = {https://doi.org/10.1016/j.cpc.2025.109611},
url = {https://www.sciencedirect.com/science/article/pii/S0010465525001134},
author = {A.Y. Pankin and J. Breslau and M. Gorelenkova and R. Andre and B. Grierson and J. Sachdev and M. Goliyad and G. Perumpilly},
}

@article{Lee_2021_triassic,
doi = {10.1088/1741-4326/ac1690},
url = {https://doi.org/10.1088/1741-4326/ac1690},
year = {2021},
month = {aug},
publisher = {IOP Publishing},
volume = {61},
number = {9},
pages = {096020},
author = {Lee, C.Y. and Seo, J. and Park, S.J. and Lee, J.G. and Kim, S.K. and Kim, B. and Byun, C.S. and Lee, Y.S. and Gwak, J.W. and Kang, J. and Jung, L. and Kim, H.-S. and Hong, S.-H. and Na, Yong-Su},
title = {Development of integrated suite of codes and its validation on KSTAR},
journal = {Nuclear Fusion},
}

@Article{Degrave2022_nature,
author={Degrave, Jonas
and Felici, Federico
and Buchli, Jonas
and Neunert, Michael
and Tracey, Brendan
and Carpanese, Francesco
and Ewalds, Timo
and Hafner, Roland
and Abdolmaleki, Abbas
and de las Casas, Diego
and Donner, Craig
and Fritz, Leslie
and Galperti, Cristian
and Huber, Andrea
and Keeling, James
and Tsimpoukelli, Maria
and Kay, Jackie
and Merle, Antoine
and Moret, Jean-Marc
and Noury, Seb
and Pesamosca, Federico
and Pfau, David
and Sauter, Olivier
and Sommariva, Cristian
and Coda, Stefano
and Duval, Basil
and Fasoli, Ambrogio
and Kohli, Pushmeet
and Kavukcuoglu, Koray
and Hassabis, Demis
and Riedmiller, Martin},
title={Magnetic control of tokamak plasmas through deep reinforcement learning},
journal={Nature},
year={2022},
month={Feb},
day={01},
volume={602},
number={7897},
pages={414-419},
issn={1476-4687},
doi={10.1038/s41586-021-04301-9},
url={https://doi.org/10.1038/s41586-021-04301-9}
}

@Article{Seo2024_nature,
author={Seo, Jaemin
and Kim, SangKyeun
and Jalalvand, Azarakhsh
and Conlin, Rory
and Rothstein, Andrew
and Abbate, Joseph
and Erickson, Keith
and Wai, Josiah
and Shousha, Ricardo
and Kolemen, Egemen},
title={Avoiding fusion plasma tearing instability with deep reinforcement learning},
journal={Nature},
year={2024},
month={Feb},
day={01},
volume={626},
number={8000},
pages={746-751},
issn={1476-4687},
doi={10.1038/s41586-024-07024-9},
url={https://doi.org/10.1038/s41586-024-07024-9}
}

@article{Abbate_2023_control, title={A general infrastructure for data-driven control design and implementation in tokamaks}, volume={89}, DOI={10.1017/S0022377822001040}, number={1}, journal={Journal of Plasma Physics}, author={Abbate, Joseph and Conlin, Rory and Shousha, Ricardo and Erickson, Keith and Kolemen, Egemen}, year={2023}, pages={895890102}}

@article{Abbate_2025_control,
doi = {10.1088/1741-4326/adc283},
url = {https://doi.org/10.1088/1741-4326/adc283},
year = {2025},
month = {apr},
publisher = {IOP Publishing},
volume = {65},
number = {5},
pages = {056014},
author = {Abbate, J. and Fable, E. and Tardini, G. and Fischer, R. and Kolemen, E. and the ASDEX Upgrade Team},
title = {Combining physics-based and data-driven models for quantitatively accurate plasma profile prediction that extrapolates well; with application to DIII-D, AUG, and ITER tokamaks},
journal = {Nuclear Fusion},
}

@article{Rothstein_2026_control,
doi = {10.1088/1741-4326/ae7f9d},
url = {https://doi.org/10.1088/1741-4326/ae7f9d},
year = {2026},
month = {jul},
publisher = {IOP Publishing},
volume = {66},
number = {7},
pages = {076050},
author = {Rothstein, A. and Farre-Kaga, H.J. and Butt, J. and Shousha, R. and Erickson, K. and Wakatsuki, T. and Steiner, P. and Kim, S.K. and Jalalvand, A. and Kolemen, E.},
title = {Enabling integrated AI control on DIII-D: a control system design with state-of-the-art experiments},
journal = {Nuclear Fusion},
}

@Article{Wang2025_disruption,
author={Wang, Allen M.
and Rea, Cristina
and So, Oswin
and Dawson, Charles
and Garnier, Darren T.
and Fan, Chuchu},
title={Active ramp-down control and trajectory design for tokamaks with neural differential equations and reinforcement learning},
journal={Communications Physics},
year={2025},
month={Jun},
day={04},
volume={8},
number={1},
pages={231},
issn={2399-3650},
doi={10.1038/s42005-025-02146-6},
url={https://doi.org/10.1038/s42005-025-02146-6}
}

@article{Yang_2025_disruption,
doi = {10.1088/1741-4326/ada396},
url = {https://doi.org/10.1088/1741-4326/ada396},
year = {2025},
month = {jan},
publisher = {IOP Publishing},
volume = {65},
number = {2},
pages = {026030},
author = {Yang, Zongyu and Zhong, Wulyu and Xia, Fan and Gao, Zhe and Zhu, Xiaobo and Li, Jiyuan and Hu, Liwen and Xu, Zhaohe and Li, Da and Zheng, Guohui and Chen, Yihang and Zhang, Junzhao and Li, Bo and Zhang, Xiaolong and Zhu, Yiren and Tong, Ruihai and Dong, Yunbo and Zhang, Yipo and Yuan, Boda and Yu, Xin and He, Zongyuhui and Tian, Wenjing and Gong, Xinwen and Xu, Min},
title = {Implementing deep learning-based disruption prediction in a drifting data environment of new tokamak: HL-3},
journal = {Nuclear Fusion},
}

@ARTICLE{Newbury_2024_differentiable_sim,
  author={Newbury, Rhys and Collins, Jack and He, Kerry and Pan, Jiahe and Posner, Ingmar and Howard, David and Cosgun, Akansel},
  journal={IEEE Access}, 
  title={A Review of Differentiable Simulators}, 
  year={2024},
  volume={12},
  number={},
  pages={97581-97604},
  doi={10.1109/ACCESS.2024.3425448}}

@misc{kingma2017_adam,
      title={Adam: A Method for Stochastic Optimization}, 
      author={Diederik P. Kingma and Jimmy Ba},
      year={2017},
      eprint={1412.6980},
      archivePrefix={arXiv},
      primaryClass={cs.LG},
      url={https://arxiv.org/abs/1412.6980}, 
}

@misc{citrin2024_torax,
      title={TORAX: A Fast and Differentiable Tokamak Transport Simulator in JAX}, 
      author={Jonathan Citrin and Ian Goodfellow and Akhil Raju and Jeremy Chen and Jonas Degrave and Craig Donner and Federico Felici and Philippe Hamel and Andrea Huber and Dmitry Nikulin and David Pfau and Brendan Tracey and Martin Riedmiller and Pushmeet Kohli},
      year={2024},
      eprint={2406.06718},
      archivePrefix={arXiv},
      primaryClass={physics.plasm-ph},
      url={https://arxiv.org/abs/2406.06718}, 
}

@inproceedings{NEURIPS2020_jax_differentiable_sim,
 author = {Schoenholz, Samuel and Cubuk, Ekin Dogus},
 booktitle = {Advances in Neural Information Processing Systems},
 editor = {H. Larochelle and M. Ranzato and R. Hadsell and M.F. Balcan and H. Lin},
 pages = {11428--11441},
 publisher = {Curran Associates, Inc.},
 title = {JAX MD: A Framework for Differentiable Physics},
 url = {https://proceedings.neurips.cc/paper\_files/paper/2020/file/83d3d4b6c9579515e1679aca8cbc8033-Paper.pdf},
 volume = {33},
 year = {2020}
}

@article{RAISSI2019_pinn,
title = {Physics-informed neural networks: A deep learning framework for solving forward and inverse problems involving nonlinear partial differential equations},
journal = {Journal of Computational Physics},
volume = {378},
pages = {686-707},
year = {2019},
issn = {0021-9991},
doi = {https://doi.org/10.1016/j.jcp.2018.10.045},
url = {https://www.sciencedirect.com/science/article/pii/S0021999118307125},
author = {M. Raissi and P. Perdikaris and G.E. Karniadakis},
}

@Article{Seo2024_pinn_optim,
author={Seo, Jaemin},
title={Solving real-world optimization tasks using physics-informed neural computing},
journal={Scientific Reports},
year={2024},
month={Jan},
day={08},
volume={14},
number={1},
pages={202},
issn={2045-2322},
doi={10.1038/s41598-023-49977-3},
url={https://doi.org/10.1038/s41598-023-49977-3}
}

@Article{Azizzadenesheli2024_neural_operator,
author={Azizzadenesheli, Kamyar
and Kovachki, Nikola
and Li, Zongyi
and Liu-Schiaffini, Miguel
and Kossaifi, Jean
and Anandkumar, Anima},
title={Neural operators for accelerating scientific simulations and design},
journal={Nature Reviews Physics},
year={2024},
month={May},
day={01},
volume={6},
number={5},
pages={320-328},
issn={2522-5820},
doi={10.1038/s42254-024-00712-5},
url={https://doi.org/10.1038/s42254-024-00712-5}
}

@article{SEO2024_net_pinn,
title = {Leveraging physics-informed neural computing for transport simulations of nuclear fusion plasmas},
journal = {Nuclear Engineering and Technology},
volume = {56},
number = {12},
pages = {5396-5404},
year = {2024},
issn = {1738-5733},
doi = {https://doi.org/10.1016/j.net.2024.07.048},
url = {https://www.sciencedirect.com/science/article/pii/S1738573324003644},
author = {J. Seo and I.H. Kim and H. Nam},
}

@article{seo2024_pre_pinn,
  title = {Past rewinding of fluid dynamics from noisy observation via physics-informed neural computing},
  author = {Seo, Jaemin},
  journal = {Phys. Rev. E},
  volume = {110},
  issue = {2},
  pages = {025302},
  numpages = {9},
  year = {2024},
  month = {Aug},
  publisher = {American Physical Society},
  doi = {10.1103/PhysRevE.110.025302},
  url = {https://link.aps.org/doi/10.1103/PhysRevE.110.025302}
}

@article{LUTJENS1996_chease,
title = {The CHEASE code for toroidal MHD equilibria},
journal = {Computer Physics Communications},
volume = {97},
number = {3},
pages = {219-260},
year = {1996},
issn = {0010-4655},
doi = {https://doi.org/10.1016/0010-4655(96)00046-X},
url = {https://www.sciencedirect.com/science/article/pii/001046559600046X},
author = {H. Lütjens and A. Bondeson and O. Sauter},
}

@article{Miller_1998,
    author = {Miller, R. L. and Chu, M. S. and Greene, J. M. and Lin-Liu, Y. R. and Waltz, R. E.},
    title = {Noncircular, finite aspect ratio, local equilibrium model},
    journal = {Physics of Plasmas},
    volume = {5},
    number = {4},
    pages = {973-978},
    year = {1998},
    month = {04},
    issn = {1070-664X},
    doi = {10.1063/1.872666},
    url = {https://doi.org/10.1063/1.872666},
}

@article{Meneghini_2017_tglfnn_eped1nn,
doi = {10.1088/1741-4326/aa7776},
url = {https://doi.org/10.1088/1741-4326/aa7776},
year = {2017},
month = {jul},
publisher = {IOP Publishing},
volume = {57},
number = {8},
pages = {086034},
author = {Meneghini, O. and Smith, S.P. and Snyder, P.B. and Staebler, G.M. and Candy, J. and Belli, E. and Lao, L. and Kostuk, M. and Luce, T. and Luda, T. and Park, J.M. and Poli, F.},
title = {Self-consistent core-pedestal transport simulations with neural network accelerated models},
journal = {Nuclear Fusion},
}

@article{qlknn_2020,
    author = {van de Plassche, K. L. and Citrin, J. and Bourdelle, C. and Camenen, Y. and Casson, F. J. and Dagnelie, V. I. and Felici, F. and Ho, A. and Van Mulders, S. and JET Contributors},
    title = {Fast modeling of turbulent transport in fusion plasmas using neural networks},
    journal = {Physics of Plasmas},
    volume = {27},
    number = {2},
    pages = {022310},
    year = {2020},
    month = {02},
    issn = {1070-664X},
    doi = {10.1063/1.5134126},
    url = {https://doi.org/10.1063/1.5134126},
}

@article{Erba_1997_bgb,
doi = {10.1088/0741-3335/39/2/004},
url = {https://doi.org/10.1088/0741-3335/39/2/004},
year = {1997},
month = {feb},
publisher = {},
volume = {39},
number = {2},
pages = {261},
author = {M Erba and A Cherubini and V V Parail and E Springmann and A Taroni},
title = {Development of a non-local model for tokamak heat transport in L-mode, H-mode and transient regimes},
journal = {Plasma Physics and Controlled Fusion},
}

@article{sauter_1999_neoclassical,
    author = {Sauter, O. and Angioni, C. and Lin-Liu, Y. R.},
    title = {Neoclassical conductivity and bootstrap current formulas for general axisymmetric equilibria and arbitrary collisionality regime},
    journal = {Physics of Plasmas},
    volume = {6},
    number = {7},
    pages = {2834-2839},
    year = {1999},
    month = {07},
    issn = {1070-664X},
    doi = {10.1063/1.873240},
    url = {https://doi.org/10.1063/1.873240},
}

@article{angioni_2000_neoclassical,
    author = {Angioni, C. and Sauter, O.},
    title = {Neoclassical transport coefficients for general axisymmetric equilibria in the banana regime},
    journal = {Physics of Plasmas},
    volume = {7},
    number = {4},
    pages = {1224-1234},
    year = {2000},
    month = {04},
    issn = {1070-664X},
    doi = {10.1063/1.873933},
    url = {https://doi.org/10.1063/1.873933},
}

@article{Martin_2008_threshold,
doi = {10.1088/1742-6596/123/1/012033},
url = {https://doi.org/10.1088/1742-6596/123/1/012033},
year = {2008},
month = {jul},
publisher = {},
volume = {123},
number = {1},
pages = {012033},
author = {Y R Martin and T Takizuka and (andthe ITPA CDBM H-mode Threshold Database Working Group)},
title = {Power requirement for accessing the H-mode in ITER},
journal = {Journal of Physics: Conference Series},
}

@article{Delabie_2026_threshold,
doi = {10.1088/1741-4326/ae39f2},
url = {https://doi.org/10.1088/1741-4326/ae39f2},
year = {2026},
month = {feb},
publisher = {IOP Publishing},
volume = {66},
number = {3},
pages = {036016},
author = {Delabie, E. and Solano, E.R. and Hughes, J.W. and Maggi, C.F. and Ryter, F. and Birkenmeier, G. and Carvalho, P. and Cavedon, M. and Chernyshova, M. and David, P. and Hillesheim, J. and Plank, U. and the EUROfusion Tokamak Exploitation Team and the ASDEX Upgrade Team and JET Contributors},
title = {Empirical scaling of the L–H threshold power for metal wall tokamaks using a multi-device database},
journal = {Nuclear Fusion},
}

@article{Onjun_2002_alpha_crit,
    author = {Onjun, T. and Bateman, G. and Kritz, A. H. and Hammett, G.},
    title = {Models for the pedestal temperature at the edge of H-mode tokamak plasmas},
    journal = {Physics of Plasmas},
    volume = {9},
    number = {12},
    pages = {5018-5030},
    year = {2002},
    month = {12},
    issn = {1070-664X},
    doi = {10.1063/1.1518474},
    url = {https://doi.org/10.1063/1.1518474},
}

@article{Cordey_2005_itpa04_scaling,
doi = {10.1088/0029-5515/45/9/007},
url = {https://doi.org/10.1088/0029-5515/45/9/007},
year = {2005},
month = {aug},
publisher = {},
volume = {45},
number = {9},
pages = {1078},
author = {Cordey, J.G. and Thomsen, K. and Chudnovskiy, A. and Kardaun, O.J.W.F. and Takizuka, T. and Snipes, J.A. and Greenwald, M. and Sugiyama, L. and Ryter, F. and Kus, A. and Stober, J. and DeBoo, J.C. and Petty, C.C. and Bracco, G. and Romanelli, M. and Cui, Z. and Liu, Y. and McDonald, D.C. and Meakins, A. and Miura, Y. and Shinohara, K. and Tsuzuki, K. and Kamada, Y. and Urano, H. and Valovic, M. and Akers, R. and Brickley, C. and Sykes, A. and Walsh, M.J. and Kaye, S.M. and Bush, C. and Hogewei, D. and Martin, Y. and Cote, A. and Pacher, G. and Ongena, J. and Imbeaux, F. and Hoang, G.T. and Lebedev, S. and Leonov, V.},
title = {Scaling of the energy confinement time with beta and collisionality approaching ITER conditions},
journal = {Nuclear Fusion},
}

@article{staebler_2007_tglf1,
    author = {Staebler, G. M. and Kinsey, J. E. and Waltz, R. E.},
    title = {A theory-based transport model with comprehensive physics},
    journal = {Physics of Plasmas},
    volume = {14},
    number = {5},
    pages = {055909},
    year = {2007},
    month = {05},
    issn = {1070-664X},
    doi = {10.1063/1.2436852},
    url = {https://doi.org/10.1063/1.2436852},
}

@article{kinsey_2008_tglf2,
    author = {Kinsey, J. E. and Staebler, G. M. and Waltz, R. E.},
    title = {The first transport code simulations using the trapped gyro-Landau-fluid model},
    journal = {Physics of Plasmas},
    volume = {15},
    number = {5},
    pages = {055908},
    year = {2008},
    month = {03},
    issn = {1070-664X},
    doi = {10.1063/1.2889008},
    url = {https://doi.org/10.1063/1.2889008},
}

@article{Citrin_2017_qualikiz,
doi = {10.1088/1361-6587/aa8aeb},
url = {https://doi.org/10.1088/1361-6587/aa8aeb},
year = {2017},
month = {nov},
publisher = {IOP Publishing},
volume = {59},
number = {12},
pages = {124005},
author = {Citrin, J and Bourdelle, C and Casson, F J and Angioni, C and Bonanomi, N and Camenen, Y and Garbet, X and Garzotti, L and Görler, T and Gürcan, O and Koechl, F and Imbeaux, F and Linder, O and Plassche, K van de and Strand, P and Szepesi, G and JET Contributors},
title = {Tractable flux-driven temperature, density, and rotation profile evolution with the quasilinear gyrokinetic transport model QuaLiKiz},
journal = {Plasma Physics and Controlled Fusion},
}

@article{Belli_2015_NEO,
doi = {10.1088/0741-3335/57/5/054012},
url = {https://doi.org/10.1088/0741-3335/57/5/054012},
year = {2015},
month = {apr},
publisher = {IOP Publishing},
volume = {57},
number = {5},
pages = {054012},
author = {Belli, E A and Candy, J},
title = {Neoclassical transport in toroidal plasmas with nonaxisymmetric flux surfaces},
journal = {Plasma Physics and Controlled Fusion},
}

@article{Snyder_2009_eped1,
    author = {Snyder, P. B. and Groebner, R. J. and Leonard, A. W. and Osborne, T. H. and Wilson, H. R.},
    title = {Development and validation of a predictive model for the pedestal height},
    journal = {Physics of Plasmas},
    volume = {16},
    number = {5},
    pages = {056118},
    year = {2009},
    month = {05},
    issn = {1070-664X},
    doi = {10.1063/1.3122146},
    url = {https://doi.org/10.1063/1.3122146},
}

@article{Lao_2022_efit,
doi = {10.1088/1361-6587/ac6fff},
url = {https://doi.org/10.1088/1361-6587/ac6fff},
year = {2022},
month = {jun},
publisher = {IOP Publishing},
volume = {64},
number = {7},
pages = {074001},
author = {Lao, L L and Kruger, S and Akcay, C and Balaprakash, P and Bechtel, T A and Howell, E and Koo, J and Leddy, J and Leinhauser, M and Liu, Y Q and Madireddy, S and McClenaghan, J and Orozco, D and Pankin, A and Schissel, D and Smith, S and Sun, X and Williams, S},
title = {Application of machine learning and artificial intelligence to extend EFIT equilibrium reconstruction},
journal = {Plasma Physics and Controlled Fusion},
}

@article{Bosch_1992_fusion_reactivity,
doi = {10.1088/0029-5515/32/4/I07},
url = {https://doi.org/10.1088/0029-5515/32/4/I07},
year = {1992},
month = {apr},
publisher = {},
volume = {32},
number = {4},
pages = {611},
author = {H.-S. Bosch and G.M. Hale},
title = {Improved formulas for fusion cross-sections and thermal reactivities},
journal = {Nuclear Fusion},
}

@article{PANKIN2004_nubeam,
title = {The tokamak Monte Carlo fast ion module NUBEAM in the National Transport Code Collaboration library},
journal = {Computer Physics Communications},
volume = {159},
number = {3},
pages = {157-184},
year = {2004},
issn = {0010-4655},
doi = {https://doi.org/10.1016/j.cpc.2003.11.002},
url = {https://www.sciencedirect.com/science/article/pii/S0010465504001109},
author = {Alexei Pankin and Douglas McCune and Robert Andre and Glenn Bateman and Arnold Kritz},
}

@article{Boyer_2019_nbi_NN1,
doi = {10.1088/1741-4326/ab0762},
url = {https://doi.org/10.1088/1741-4326/ab0762},
year = {2019},
month = {mar},
publisher = {IOP Publishing},
volume = {59},
number = {5},
pages = {056008},
author = {Boyer, M.D. and Kaye, S. and Erickson, K.},
title = {Real-time capable modeling of neutral beam injection on NSTX-U using neural networks},
journal = {Nuclear Fusion},
}

@article{MOROSOHK2021_nbi_nn2,
title = {Accelerated version of NUBEAM capabilities in DIII-D using neural networks},
journal = {Fusion Engineering and Design},
volume = {163},
pages = {112125},
year = {2021},
issn = {0920-3796},
doi = {https://doi.org/10.1016/j.fusengdes.2020.112125},
url = {https://www.sciencedirect.com/science/article/pii/S0920379620306736},
author = {Shira M. Morosohk and Mark D. Boyer and Eugenio Schuster},
}

@article{WANG2023_nbi_nn3,
title = {Neural network model of neutral beam injection in the EAST tokamak to enable fast transport simulations},
journal = {Fusion Engineering and Design},
volume = {191},
pages = {113514},
year = {2023},
issn = {0920-3796},
doi = {https://doi.org/10.1016/j.fusengdes.2023.113514},
url = {https://www.sciencedirect.com/science/article/pii/S0920379623000984},
author = {Z. Wang and S. Morosohk and T. Rafiq and E. Schuster and M.D. Boyer and W. Choi},
}

@article{Rothstein_2024_nbi_nn4,
doi = {10.1088/1741-4326/ad64e6},
url = {https://doi.org/10.1088/1741-4326/ad64e6},
year = {2024},
month = {jul},
publisher = {IOP Publishing},
volume = {64},
number = {9},
pages = {096020},
author = {Rothstein, Andrew and Jalalvand, Azarakhsh and Abbate, Joseph and Erickson, Keith and Kolemen, Egemen},
title = {Initial testing of Alfvén eigenmode feedback control with machine-learning observers on DIII-D},
journal = {Nuclear Fusion},
}

@book{huba1998nrl,
  title={NRL plasma formulary},
  author={Huba, Joseph Donald},
  volume={6790},
  number={98-358},
  year={1998},
  publisher={Naval Research Laboratory}
}

@article{Albajar_2001_synchrotron,
doi = {10.1088/0029-5515/41/6/301},
url = {https://doi.org/10.1088/0029-5515/41/6/301},
year = {2001},
month = {jun},
publisher = {},
volume = {41},
number = {6},
pages = {665},
author = {F. Albajar and J. Johner and G. Granata},
title = {Improved calculation of synchrotron radiation losses 
in realistic tokamak plasmas},
journal = {Nuclear Fusion},
}

@article{JUHN2020_density_feedback_control,
title = {Low-risk beginning of the density control on the KSTAR plasmas},
journal = {Fusion Engineering and Design},
volume = {160},
pages = {111706},
year = {2020},
issn = {0920-3796},
doi = {https://doi.org/10.1016/j.fusengdes.2020.111706},
url = {https://www.sciencedirect.com/science/article/pii/S0920379620302544},
author = {June-Woo Juhn and Y.S. Hwang and S.H. Hahn and Y.U. Nam and K.P. Kim and Y.O. Kim and J.I. Song},
}

@article{Mlynek_2011_density_feedback_control,
doi = {10.1088/0029-5515/51/4/043002},
url = {https://doi.org/10.1088/0029-5515/51/4/043002},
year = {2011},
month = {mar},
publisher = {},
volume = {51},
number = {4},
pages = {043002},
author = {Mlynek, A. and Reich, M. and Giannone, L. and Treutterer, W. and Behler, K. and Blank, H. and Buhler, A. and Cole, R. and Eixenberger, H. and Fischer, R. and Lohs, A. and Lüddecke, K. and Merkel, R. and Neu, G. and Ryter, F. and Zasche, D. and the ASDEX Upgrade Team},
title = {Real-time feedback control of the plasma density profile on ASDEX Upgrade},
journal = {Nuclear Fusion},
}

@article{Zheng_2013_density_feedback_control,
doi = {10.1088/0741-3335/55/11/115010},
url = {https://doi.org/10.1088/0741-3335/55/11/115010},
year = {2013},
month = {oct},
publisher = {IOP Publishing},
volume = {55},
number = {11},
pages = {115010},
author = {Zheng, Xingwei and Li, Jiangang and Hu, Jiansheng and Li, Jiahong and Ding, Rui and Cao, Bin and Wu, Jinhua},
title = {Comparison between gas puffing and supersonic molecular beam injection in plasma density feedback experiments in EAST},
journal = {Plasma Physics and Controlled Fusion},
}

@article{JANKY2015_density_feedback_control,
title = {Plasma density control in real-time on the COMPASS tokamak},
journal = {Fusion Engineering and Design},
volume = {96-97},
pages = {637-640},
year = {2015},
note = {Proceedings of the 28th Symposium On Fusion Technology (SOFT-28)},
issn = {0920-3796},
doi = {https://doi.org/10.1016/j.fusengdes.2015.04.065},
url = {https://www.sciencedirect.com/science/article/pii/S092037961500294X},
author = {F. Janky and M. Hron and J. Havlicek and M. Varavin and F. Zacek and J. Seidl and R. Panek},
}

@article{SHKim_2018_heating_overshoot_transition,
doi = {10.1088/1741-4326/aab034},
url = {https://doi.org/10.1088/1741-4326/aab034},
year = {2018},
month = {mar},
publisher = {IOP Publishing},
volume = {58},
number = {5},
pages = {056013},
author = {Kim, S.H. and Casper, T.A. and Snipes, J.A.},
title = {Investigation of key parameters for the development of reliable ITER baseline operation scenarios using CORSICA},
journal = {Nuclear Fusion},
}

@article{BRAY2026_pedestal_fixed,
title = {Analysis of H-mode and L-mode plasmas with a liquid Sn CPS divertor test unit in the ASDEX Upgrade tungsten divertor},
journal = {Nuclear Materials and Energy},
volume = {47},
pages = {102121},
year = {2026},
issn = {2352-1791},
doi = {https://doi.org/10.1016/j.nme.2026.102121},
url = {https://www.sciencedirect.com/science/article/pii/S2352179126000645},
author = {Elisabetta Bray and Daniel Fajardo and Clemente Angioni and Oleg Samoylov and Ralph Dux and Giuseppe F. Nallo and Chiara Marchetto and Fabio Subba},
}

@article{Schramm_2025_pedestal_fixed,
doi = {10.1088/1741-4326/adefae},
url = {https://doi.org/10.1088/1741-4326/adefae},
year = {2025},
month = {aug},
publisher = {IOP Publishing},
volume = {65},
number = {9},
pages = {096003},
author = {Schramm, Raphael and Bock, Alexander and Fable, Emiliano and Stober, Jörg and van Mulders, Simon and Maraschek, Marc and Fischer, Rainer and Burckhart, Andreas and Schneider, Philip and Sauter, Olivier and Zohm, Hartmut and Upgrade Team, the ASDEX and JET Contributors},
title = {Generalization of a fast, predictive model for tokamak scenario design in ASTRA and comparison to a RAPTOR model},
journal = {Nuclear Fusion},
}

@article{deepmind2020_jax_optax,
  title = {The {D}eep{M}ind {JAX} {E}cosystem},
  author = {DeepMind and Babuschkin, Igor and Baumli, Kate and Bell, Alison and Bhupatiraju, Surya and Bruce, Jake and Buchlovsky, Peter and Budden, David and Cai, Trevor and Clark, Aidan and Danihelka, Ivo and Dedieu, Antoine and Fantacci, Claudio and Godwin, Jonathan and Jones, Chris and Hemsley, Ross and Hennigan, Tom and Hessel, Matteo and Hou, Shaobo and Kapturowski, Steven and Keck, Thomas and Kemaev, Iurii and King, Michael and Kunesch, Markus and Martens, Lena and Merzic, Hamza and Mikulik, Vladimir and Norman, Tamara and Papamakarios, George and Quan, John and Ring, Roman and Ruiz, Francisco and Sanchez, Alvaro and Sartran, Laurent and Schneider, Rosalia and Sezener, Eren and Spencer, Stephen and Srinivasan, Srivatsan and Stanojevic, Milos and Stokowiec, Wojciech and Wang, Luyu and Zhou, Guangyao and Viola, Fabio},
  url = {http://github.com/google-deepmind},
  year = {2020},
}

@article{bkim2026_bayesian_reactor,
  title={Bayesian optimization for design space exploration of tokamak fusion reactors},
  author={Kim, Boseong and Kwon, Jae-Min and Jo, GaHyung and Hong, Bong Guen and Hahn, Sang Hee and Figuera-Michal, Darian and Yoon, Eisung},
  year={2026},
  journal={Available at SSRN 6828995}
}

@misc{jskim2024_rl_reactor,
      title={Design Optimization of Nuclear Fusion Reactor through Deep Reinforcement Learning}, 
      author={Jinsu Kim and Jaemin Seo},
      year={2024},
      eprint={2409.08231},
      archivePrefix={arXiv},
      primaryClass={physics.plasm-ph},
      url={https://arxiv.org/abs/2409.08231}, 
}

@article{NAM2025_ml_prediction,
title = {Machine learning based energy confinement time extrapolation via multi-tokamak database},
journal = {Fusion Engineering and Design},
volume = {221},
pages = {115369},
year = {2025},
issn = {0920-3796},
doi = {https://doi.org/10.1016/j.fusengdes.2025.115369},
url = {https://www.sciencedirect.com/science/article/pii/S0920379625005654},
author = {Hyungkeun Nam and Jaemin Seo},
}

@Article{Yang2026_ml_prediction,
author={Yang, Zongyu
and Yang, Zhenghao
and Tian, Wenjing
and Li, Jiyuan
and Sun, Xiang
and Zheng, Guohui
and Liu, Songfen
and Wu, Niannian
and Li, Rongpeng
and Xu, Zhaohe
and Li, Bo
and Shi, Zhongbing
and Gao, Zhe
and Chen, Wei
and Ji, Xiaoquan
and Xu, Min
and Zhong, Wulyu},
title={FusionMAE, a self-supervised pretrained model to optimize and simplify diagnostic and control of fusion plasma},
journal={Communications Physics},
year={2026},
month={May},
day={01},
issn={2399-3650},
doi={10.1038/s42005-026-02626-3},
url={https://doi.org/10.1038/s42005-026-02626-3}
}

@article{Abadi2026_tensorflow,
author = {Abadi, Martin},
title = {TensorFlow: learning functions at scale},
year = {2016},
issue_date = {September 2016},
publisher = {Association for Computing Machinery},
address = {New York, NY, USA},
volume = {51},
number = {9},
issn = {0362-1340},
url = {https://doi.org/10.1145/3022670.2976746},
doi = {10.1145/3022670.2976746},
journal = {SIGPLAN Not.},
month = sep,
pages = {1},
numpages = {1}
}

@inproceedings{NEURIPS2019_pytorch,
 author = {Paszke, Adam and Gross, Sam and Massa, Francisco and Lerer, Adam and Bradbury, James and Chanan, Gregory and Killeen, Trevor and Lin, Zeming and Gimelshein, Natalia and Antiga, Luca and Desmaison, Alban and Kopf, Andreas and Yang, Edward and DeVito, Zachary and Raison, Martin and Tejani, Alykhan and Chilamkurthy, Sasank and Steiner, Benoit and Fang, Lu and Bai, Junjie and Chintala, Soumith},
 booktitle = {Advances in Neural Information Processing Systems},
 pages = {1},
 publisher = {Curran Associates, Inc.},
 title = {PyTorch: An Imperative Style, High-Performance Deep Learning Library},
 url = {https://proceedings.neurips.cc/paper\_files/paper/2019/file/bdbca288fee7f92f2bfa9f7012727740-Paper.pdf},
 volume = {32},
 year = {2019}
}


\end{document}